\documentclass{article}	
\usepackage{multicol}
\usepackage{graphicx}
\usepackage{color}
\usepackage{float}
\usepackage{amsmath}
\usepackage[margin=0.5in]{geometry}
\usepackage{cite}
\usepackage{MnSymbol}
\usepackage{authblk}
\definecolor{mycolor1}{rgb}{0.3, 0.9, 0.9}
\definecolor{mycolor2}{rgb}{0.1, 0.7, 0.3}
\newcommand{\subfigimg}[3][,]{%
  \setbox1=\hbox{\includegraphics[#1]{#3}}% Store image in box
  \leavevmode\rlap{\usebox1}% Print image
  \rlap{\hspace*{10pt}\raisebox{\dimexpr\ht1-1\baselineskip}{#2}}% Print label
  \phantom{\usebox1}% Insert appropriate spcing
}
\begin{document}
\title{Enhanced plasmonic coloring of silver and formation of large laser-induced periodic surface structures using multi-burst picosecond pulses}

%% Notice placement of commas and superscripts and use of &
%% in the author list
\author[1,2,*]{J.-M. Guay}
\author[1,2]{A. Cal\`a Lesina}
\author[1]{J. Baxter}
\author[1]{M. Charron}
\author[1,2]{G. C\^ot\'e}
\author[1,2]{L. Ramunno}
\author[1,2,4]{P. Berini}
\author[1,2,3]{A. Weck}

\affil[1]{Department of Physics, University of Ottawa, ON, K1N 6N5, Canada}
\affil[2]{Centre for Research in Photonics, University of Ottawa, ON, K1N 6N5, Canada}
\affil[3]{Department of Mechanical Engineering, University of Ottawa, ON, K1N 6N5, Canada}
\affil[4]{School of Electrical Engineering and Computer Science, University of Ottawa, ON, K1N 6N5, Canada}

\affil[*]{Corresponding author: jguay036@uottawa.ca}
\maketitle

\begin{abstract}
We report on the creation of angle-independent colors on silver using closely time-spaced laser bursts. The use of burst mode, compared to traditional non-burst is shown to increase the Chroma (color saturation) by $\sim$50$\%$ and to broaden the lightness range by up to $\sim$60$\%$. Scanning electron microscope analysis of the surfaces created using burst mode, reveal the creation of 3 distinct sets of laser induced periodic surface structures (LIPSS): low spatial frequency LIPSS (LSFL), high spatial frequency LIPSS (HSFL) and large laser-induced periodic surface structures (LLIPSS) that are 10 times the laser wavelength and parallel to the laser polarization. Nanoparticles are responsible for each plasmonic color and their distributions are observed to be similar for both burst and non-burst modes, indicating that the underlying structures (\textit{i.e.} LIPSSs) are responsible for the increased Chroma and Lightness. Two-temperature model simulations of silver irradiated by laser bursts show significant increase in the electron-phonon coupling coefficient which is crucial for the creation of well-defined ripple structures. Finite difference time domain (FDTD) simulations of the colored surfaces show the increase in Chroma to be attributable to the HSFL arising in burst mode.
\end{abstract}

\begin{multicols}{2}
\section{Introduction}
Metallic nanoparticles (NPs) exhibit unique plasmonic responses when exposed to optical radiation, leading to promising applications in many fields, such as material processing \cite{Vorobyev2013a, Fan2014, Guay2016a}, solar cells \cite{Salvador2012,Rai2015} and medicine \cite{Huang2011,Catchpole2008}. It was shown that the surface plasmon (SP) resonance of nanoparticles is influenced by factors such as size, volume fraction, shape and permittivity \cite{Mie,Doyle1989,Murray2007}. Coloring metal surfaces by exploiting surface plasmon resonances is an emerging topic and multiple fabrication techniques exist to tune the plasmonic response on the surface of metals \cite{Gallinet2015,Roberts2014,Tan2014,Cheng2015,Clausen2014}. However, current processes are tedious and time consuming. The use of lasers for the coloring of metals is still a relatively new concept. The coloring of metals using lasers has been described by a three step process: (1) laser induced melting, (2) spallation and (3) redeposition and fusion of NPs onto the substrate\cite{Vorobyev2013a}.  The exposure of metals to laser light can also produce laser-induced periodic surface structures (LIPSS). Their formation is understood as resulting from the interference between the laser light and a surface scattered wave \cite{Huang2009,Fauchet1982,Sipe1983,phenon}. Recently, it has been suggested that excited surface plasmon polaritons (SPPs) during laser irradiation could play a crucial role in the formation of LIPSS and that surface plasmon (SP)-grating assisted coupling would be responsible for the creation of laser-induced periodic surface structures (LIPSS) with periods smaller than the wavelength of light \cite{Huang2009}. Such ripples are typically oriented perpendicular to the light polarization and governed by the transverse magnetic (TM) characteristics of excited SPs \cite{Huang2009}. Periodic structures with periodicity of $\lambda$/10 called high spatial frequency LIPSS (i.e. HSFL) were also observed, however, their origin is still debated \cite{Yao2012a,Bonse2011}.\\

Recent advancements in laser burst technology, consisting of closely time-spaced laser pulses, have found many applications in the machining of glass and metals \cite{Hartmann2007,Herman1999}. The use of burst in the ablation of metals was shown to increase the ablation rate due to interactions of subsequent pulses within the burst with the already hot surface \cite{Ren2013,Hu2009}.
Typical two temperature models are known to neglect the temperature dependence of many parameters \cite{Ren2013}, a valid assumption when modeling light-matter interaction with a long pulse separation. However, in burst mode, a modified two-temperature model is needed in order to consider the temperature dependence of the interaction. Hu \textit{et al.} \cite{Hu2009} showed that the use of burst with picosecond pulse separation significantly decreased the electron relaxation time and increased the electron-phonon coupling coefficients of copper by a factor of 11 due to the higher electron temperature perceived by the next pulse within the burst. \begin{figure}[H]
  \centering
  \begin{tabular}{@{}p{.5\linewidth}@{}}
		\subfigimg[width=\linewidth]{\textbf{\textcolor{black}{a)}}}{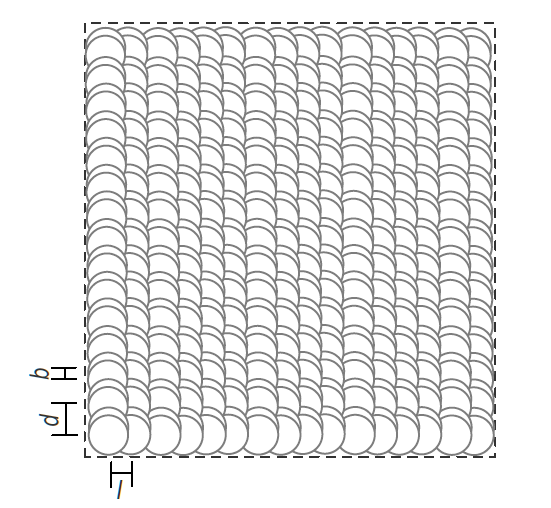}\\
				\subfigimg[width=\linewidth]{\textbf{\textcolor{black}{b)}}}{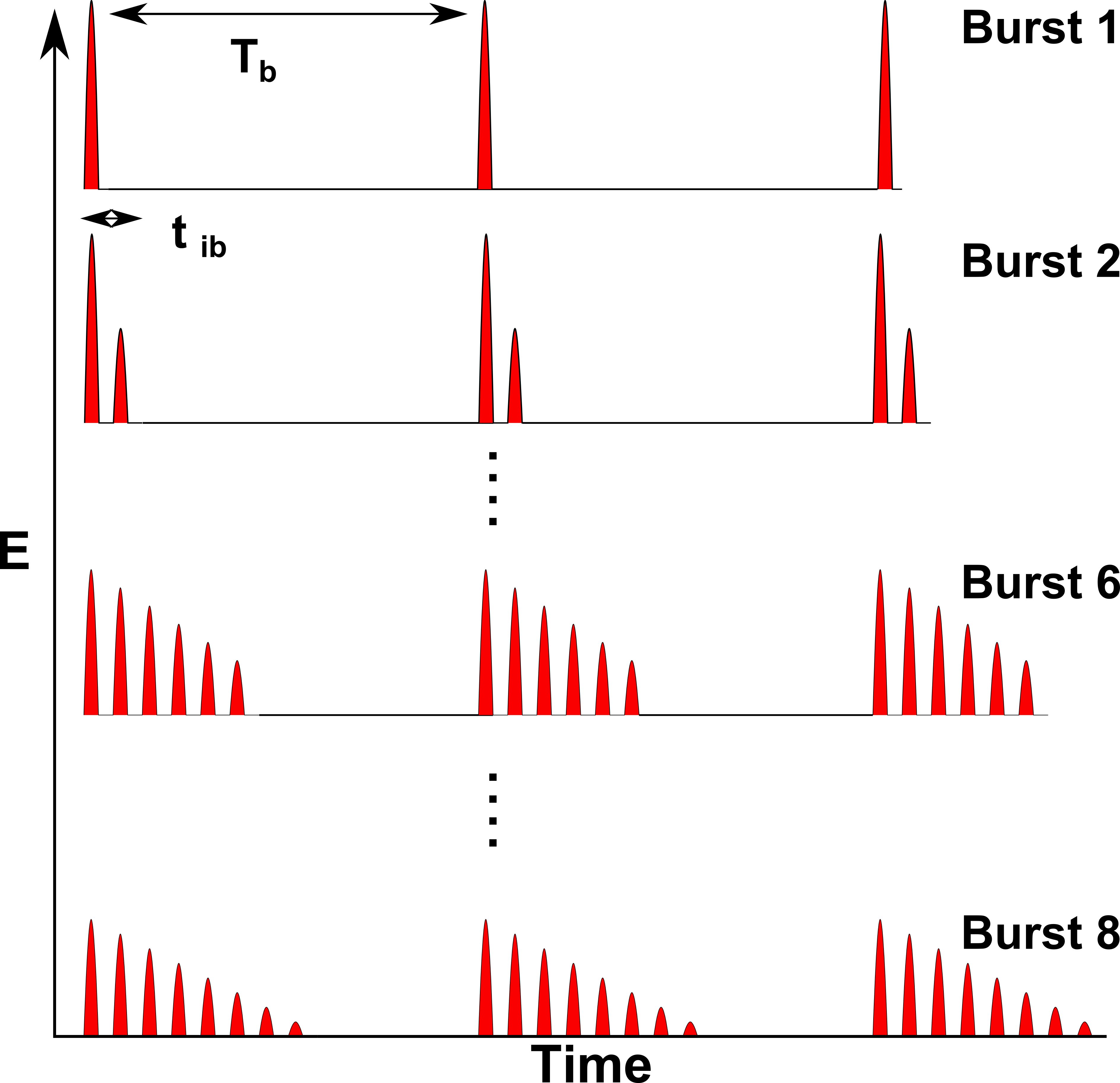}

	 \end{tabular}
\caption{(a) Schematic of the coloring process for coloring silver. The raster scanning of the surface is done in a top to bottom fashion where $l$ is the spacing between each successive laser line, d is the distance between successive laser shots and b is the distance between the pulses within the bursts. (b) Schematic of several bursts; the burst number represents the number of pulses separated by $t_{ib}$=12.8ns produced by the laser at a given repetition frequency, $f_b = 1/T_b$ where $T_b$ is the burst period.}
\label{fig1}
\end{figure}
Electron-phonon coupling plays a crucial role in the creation of LIPSS. An increase in electron-phonon coupling could have major implication in the creation of LIPSS on metals typically known not to respond well to laser irradiation, such as silver and gold \cite{Wang2005}. In addition, burst was shown to interact with the initial ablation plume causing the plume to augment significantly in volume \cite{Hartmann2007}, redepositing nanoparticles onto a larger area.
\\
In this paper, we report on the coloring of silver using burst mode. The formation of 3 distinct of LIPSS: 1) low spatial frequency LIPSS (LSFL), 2) high spatial frequency LIPSS (HSFL) and 3) large LIPSS (LLIPSS) is presented. The use of burst on silver is shown to increase the Chroma values by $\sim$50$\%$ and extend the lightness range by up to $\sim$60$\%$. Comparisons between colored surfaces produced with burst and non-burst reveal the same statistical distribution of nanoparticles, pointing to the underlying LIPSS structure as likely responsible for the increased Chroma and Lightness range. Two-temperature model simulations of the burst process shows that each pulse within a burst impinges on an increasingly hot surface and also reveals an increase in the electron-phonon coupling responsible for the well-defined LIPSS. Finite-difference time-domain (FDTD) simulations of the surfaces reveal the HSFL to be responsible for the increase Chroma values observed experimentally.

\section{Methods}

The irradiation of silver surfaces was carried out in a raster-scanning fashion, as sketched in Figure 1a, with 1064 nm light from a Duetto mode-locked laser (Nd:YVO$_{4}$, Time-Bandwidth Product) producing 10 ps pulses. The burst repetition rate $f_b = 1/T_b$ ($T_b$ is the burst period) was tuneable from 50 to 8200 kHz. The time separation between each pulse within a burst is $t_{ib}$ = 12.8 ns, and is fixed, Figure 1(b). A selection of 1 to 8 pulses within each burst was available. The distribution of energy within each shot was controlled using software provided by the manufacturer (FlexBurst$^{TM}$). For this paper, the energy of the burst pulses are such that the amplitude of each pulse within a burst is decreasing (i.e. reducing in energy). The light was focused on the silver surface using an F-theta lens (f = 163 mm, Rodenstock). The laser was fully electronically integrated and enclosed by a third party for industrial applications (GPC-PSL, FOBA). For accurate focusing, the surface of the samples was located using a touch probe system. The silver samples were of 99.99$\%$ purity. Samples were not polished prior to machining to meet requirements of reproducibility in industrial application. \\
For machining, the samples were placed on a 3-axis stage with a resolution of 1 $\mu m$ in both the lateral and axial directions. The samples were raster scanned using galvanometric XY mirrors (Turboscan 10, Raylase) displacing the beam in a top to bottom fashion, Figure 1(a). The laser polarization was kept parallel to the laser machining direction. The laser power was computer-controlled via a laser interface and calibrated using a power meter (3A-P-QUAD, OPHIR). A Gaussian beam radius of $\approx$ 14 $\mu m$ was obtained from a semi-logarithmic plot of the square diameter of the modified region, measured with a scanning electron microscope (SEM), as a function of laser pulse energy following the method described in \cite{Jandeleit1996}. High resolution SEM (JSM-7500F FESEM, JEOL) images were obtained using secondary electron imaging (SEI) mode. Colors were quantified using a Konica Minolta CR-241 Chroma meter in the CIELCH color space, 2 observer and illuminant C (North sky daylight).\\
\\
Two-Temperature model (TTM) simulations were done using parallel computing and implemented on CUDA allowing for the use of a GPU. The TTM model equations were solved using the finite-difference method using a space discretisation step size of 2.5 nm \cite{Kirkwood2007}. The time step was chosen to be on the order of attoseconds for numerical stability.
Three dimensional finite-difference time-domain simulations \cite{Taflove2005,Taflove2013} were performed to determine the origin of the Chroma increase. We used in-house 3D-FDTD parallel code \cite{Lesina2015, Vaccari2014} on an IBM BlueGene/Q supercomputer (64k cores) part of the Southern Ontario Smart Computing Innovation Platform (SOSCIP). The dispersive Drude-2CP model for silver was used. The simulation domain was discretized with a space step of 0.25 nm.
%The dispersive Drude+2CP model for silver was used. The simulation domain was discretized with a space step of 0.25 nm. The simulations ran on the IBM BlueGene/Q of the Southern Ontario Smart Computation Innovation Platform (SOSCIP) using up to 16k cores. 

\section{Results and Discussion}
Figure 2 demonstrates the angle-independent coloring of silver coins using burst. The different colors were obtained by varying the laser fluence, marking speed, line spacing and number of pulses per burst.
%A burst of 5 with a total fluence of 4.06 $J/cm^{2}$ marking speed of 44 mm/s, and different line spacing were used to create the (a) blue (11$\mu m$), (b) red (15 $\mu m$) and (c) purple colors (13 $\mu m$).
In previous work, we showed that each color was bound to a total accumulated fluence curve and that by simply changing the line spacing between successive lines one could generate a full palette of colors \cite{Guay2016a}. Figure 2 (d-e) shows the colors achieved by keeping the laser parameters fixed and simply changing the number of pulses per burst (column) and line spacing (row). Different colors are obtained with a different number of pulses per burst for the same line spacing. The highest Chroma values were generally obtained with the highest number of pulses per burst. 
\begin{figure}[H]
  \centering
  %\begin{tabular}{@{}p{.33\linewidth}@{}p{.33\linewidth}@{}p{.33\linewidth}@{}}
	%\subfigimg[width=\linewidth]{\textbf{\textcolor{white}{a)}}}{blue.JPG}&
		%\subfigimg[width=\linewidth]{\textbf{\textcolor{white}{b)}}}{red.JPG}&
			%\subfigimg[width=\linewidth]{\textbf{\textcolor{white}{c)}}}{purple.JPG}
  %\end{tabular}
	%\begin{tabular}{@{}p{.55\linewidth}@{}p{.445\linewidth}@{}}
		%\subfigimg[width=\linewidth]{\textbf{\textcolor{white}{(d)}}}{burst2.JPG}&
				%\subfigimg[width=\linewidth]{\textbf{\textcolor{white}{(e)}}}{burst1.JPG}
	 %\end{tabular}
  \begin{tabular}{@{}p{1\linewidth}@{}}
		\subfigimg[width=\linewidth]{\textbf{\textcolor{white}{}}}{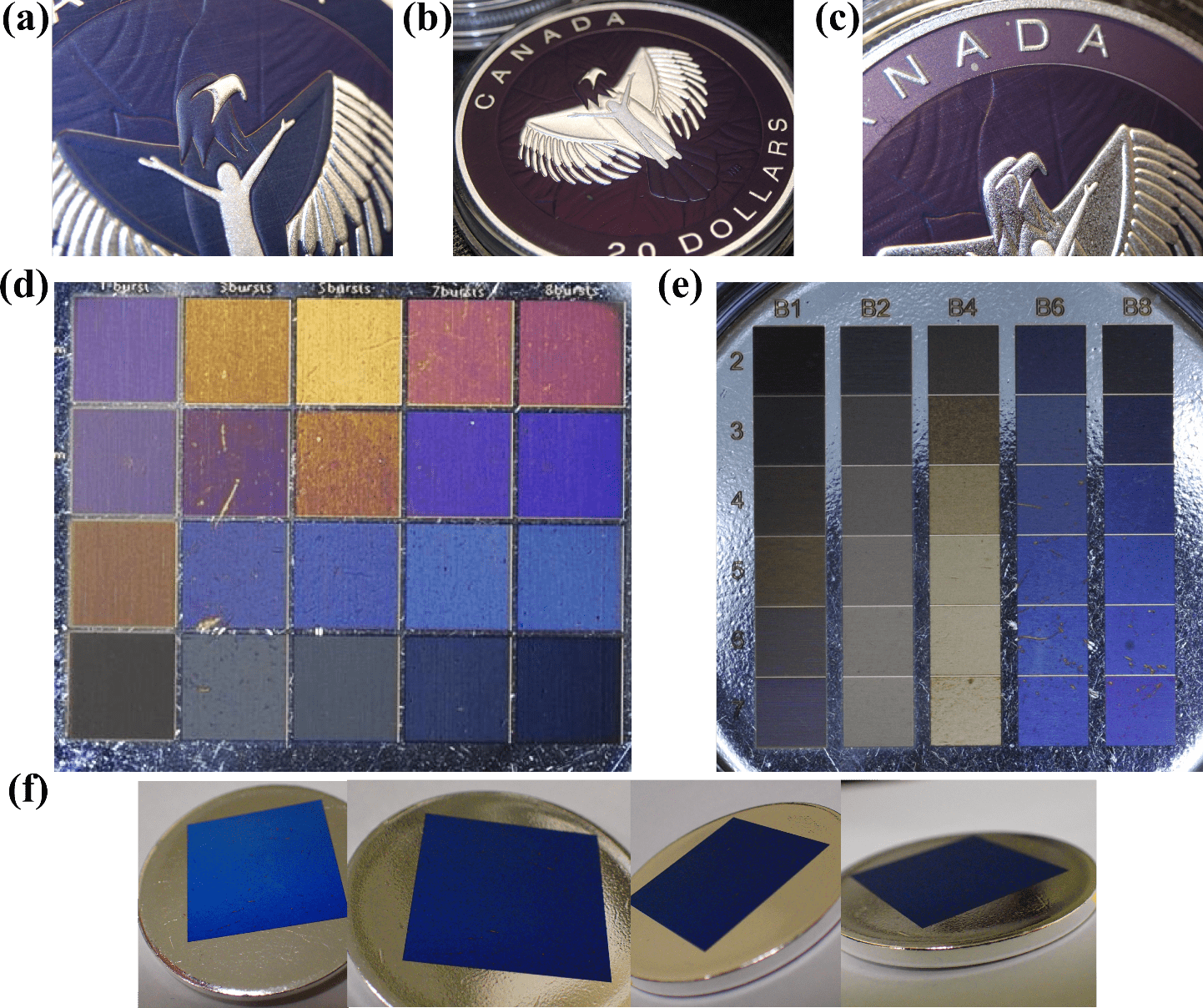}
	 \end{tabular}
\caption{(a-c) Photographs of topographical colored silver eagles. The (a) blue, (b) red and (c) purple colors were obtained by varying the line spacing between each successive line using a marking speed of 44 mm/s and a laser fluence of 4.06 $J/cm^{2}$. Photograph of colors obtained on silver comparing burst and non-burst for a total laser fluence of 15.88 J/cm$^2$ with a marking speed of (d) 50 mm/s and (e) 100 mm/s. From top to bottom (rows) the line spacing is increasing, and from left to right (columns), the number of pulses per burst is (d) 1,3,5,7 and 8, and (e) 1,2,4,6 and 8. (f) Photograph of colored of flat silver surface observed at various angles; color was made with a fluence of 15.88 J/cm$^2$, marking speed of 100 mm/s, line spacing of 6 $\mu m$ and a burst of 8 pulses.}
\label{fig2}
\end{figure}
The total energy deposited on the surface remains the same, however, the division of the laser energy into multiple closely spaced pulses (within a burst), Figure 1 (b), is found to have significant effects on the colors. The colors with non-burst (one pulse per burst) were visually less desirable and appeared mat or even burnt. We make the valid assumption that each of the pulses within the burst overlap perfectly on the surface given the timescale and the relative motion of the galvanometric mirrors.

\begin{figure}[H]
  \centering
  \begin{tabular}{@{}p{1\linewidth}@{}}
		\subfigimg[width=\linewidth]{\textbf{\textcolor{black}{(a)}}}{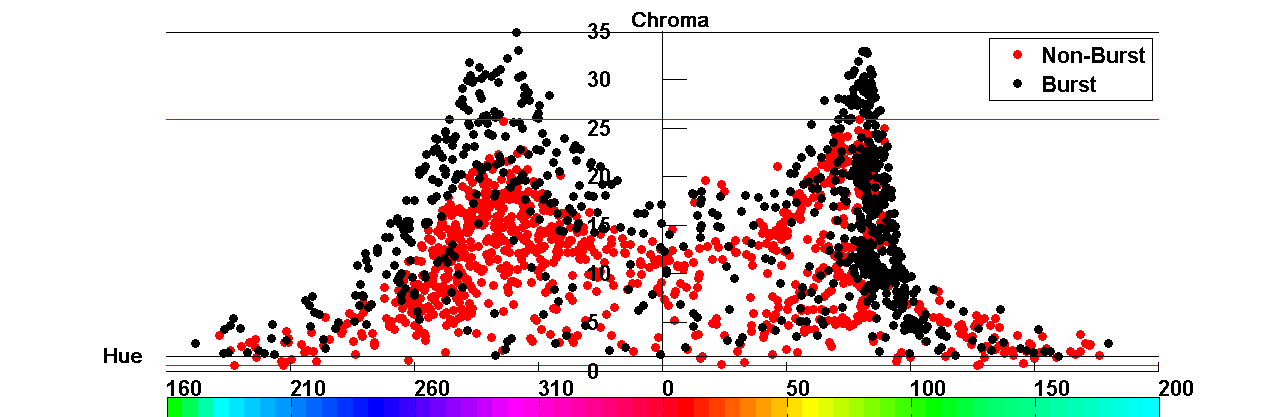}\\
				\subfigimg[width=\linewidth]{\textbf{\textcolor{black}{(b)}}}{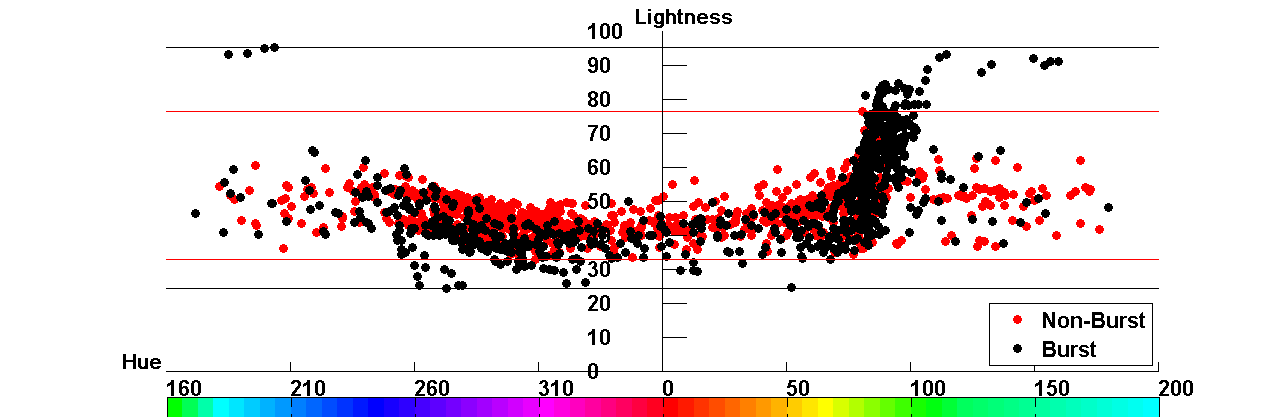}\\
	 \end{tabular}
	 \begin{tabular}{@{}p{.8\linewidth}@{}}
	\subfigimg[width=\linewidth]{\textbf{\textcolor{black}{(c)}}}{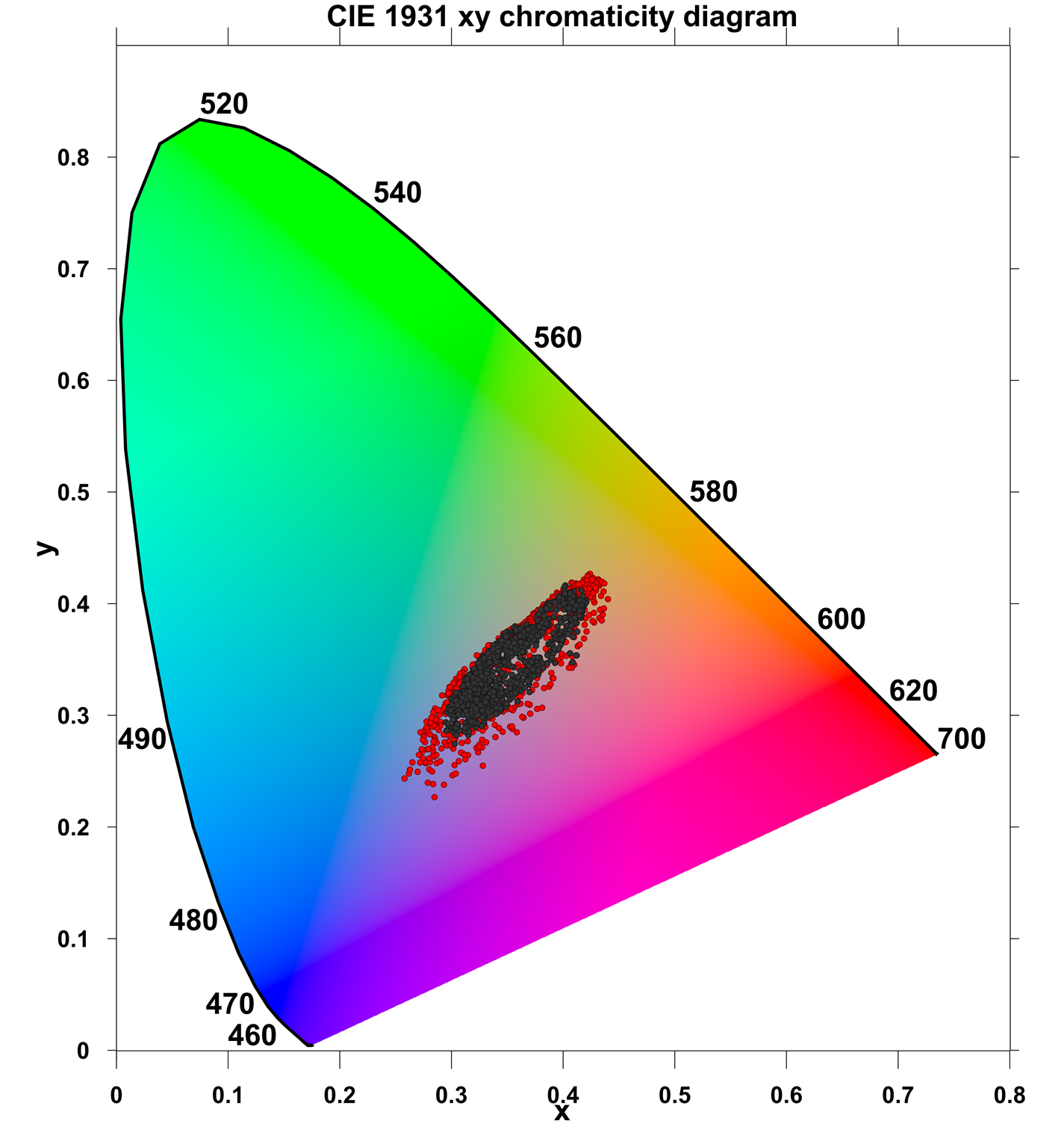}
	 \end{tabular}
\caption{Graph comparing (a) Chroma and (b) Lightness versus Hue for burst and non-burst. The red and black horizontal lines delimit the ranges of non-burst and burst, respectively. Panel (c) is a CIE diagram representation of the burst (red dots) and non-burst (black dots) colors.}
\label{fig3}
\end{figure}
%We make the valid assumption that the spatial separation between the pulses within the burst arriving on the substrate is negligible compared to the relative motion of the mirrors.  

 Interestingly, in the case of burst, a significant increase in Chroma ($\sim$50$\%$) can be observed over the full range of colors, with the exception of yellow, Figure 3(a). The range of lightness for burst is also significantly larger than for non-burst, by 60$\%$ for particular Hue values, Figure 3(b). The increased range of Lightness has far-reaching implications for the coloring of silver because it allows for a significantly broader palette of colors (\textit{e.g.} sky blue or navy blue). Figure 3 (c) is a CIE diagram comparing the colors obtained via burst and non-burst where again the larger range of colors obtained by burst can be observed.
\\
\\
In a previous publication we explained the origin of the colors for non-burst as a plasmonic effect coming from the coupling of small and medium nanoparticles on the surface \cite{Guay2016a}. SEM image analysis and FDTD simulations of colored surfaces showed that different colors originated from different particle arrangements on the surface of silver due to their collective plasmonic oscillations. The surface of the colored regions of silver were relatively flat with no distinct elevation. In contrast, SEM images of the burst surfaces shown in Fig. 4, reveal surprising structures that are not present in the non-burst case. In addition to LIPSS, other much larger periodic structures (\textit{i.e.} LLIPSS) are observed that are 10 times the wavelength of the incoming laser light and are oriented along the direction of the laser polarization. At normal laser incidence, their creation cannot be explained by light-scattering interference or light-plasmon interference theories \cite{Huang2009,Fauchet1982,Sipe1983,phenon}.
\\
\\ 
The structures appear only under burst and are more pronounced for a burst of 2 pulses due in part to the higher energy per pulse within the burst. The evolution of the structures with decreasing number of pulses per burst (\textit{i.e.} higher energy per pulse) gives us insight into the creation of the LLIPSS structures. For 8 pulse bursts, Figure 4 (a), random hole-like structures are seen on the surface. With a decreasing number of pulses per burst (b,c) the number of random hole-like structures increases and they eventually coalesce (d) to form structures that are 10 times the wavelength of light. This indicates not only a feedback mechanism in the process but possibly dynamic interaction with the molten surface. Close-up views of the surfaces (e-h) reveal well-defined periodic structures that are parallel to the laser polarization with a periodicity of $\approx$ $\lambda$/10 (\textit{i.e.} HSFL).
%An observation that indicates not only a feedback mechanism in the process but also a dynamic interaction with the molten surface simply from the absence of such structures in the non-burst regime.
%Chroma values with (i.e. 2 to 8 burst) and without (i.e. burst of 1) multi-burst strikingly different as seen in Figure 3 (a). 
\\
\\
High-resolution SEM images of surfaces revealed the presence of 3 particle classes \cite{Guay2016a}. Measurements and comparisons of the number of particles for the different colors obtained with and without burst are shown in Figure 5. The number of small nanoparticles ($\sim$ 6.5 nm radius) per unit area is similar for a given color. Previous work showed the small NPs to play a dominant role in the plasmonic coloring of silver \cite{Guay2016a}. From this observation, the differences in Chroma and Lightness can only come from differences in the underlying surface structure. Huang \textit{et al.}\cite{Huang2009} showed that crevasses create stronger electric fields than non-topographical surfaces when exposed to optical radiation. This would serve to explain the enhanced plasmon resonance of a particle on the edge of a crevasse, consequently increasing absorption. This could in turn cause the Chroma to increase significantly.
\\
\subsection{Two-Temperature Model}
The TTM is a 1-D temperature diffusion model representing the diffusion of the laser energy in the electron and lattice subsystems \cite{Chen2001}. The relations describing the dynamics of diffusion is given by coupled differential equations where the laser energy is diffused mainly through the electron system. The energy is transferred to the lattice via collisions which are represented by electron-phonon coupling. The relations are:
\begin{equation}
C_{e}\frac{\delta T_e}{\delta t}=\frac{\delta}{\delta z}(k_e\frac{\delta T_e}{\delta z})-g(T_e -T_l)+S(z,t)
\end{equation}
\begin{equation}
C_{l}\frac{\delta T_l}{\delta t}=\frac{\delta}{\delta z}(k_l\frac{\delta T_l}{\delta z})+g(T_e -T_l)
\end{equation}
where $C_e$ and $C_l$ are the electronic and lattice thermal heat capacities respectively, $k_e$ is the electronic thermal conductivity, $k_l$ is the lattice thermal conductivity, g is the electron-phonon coupling, S is the energy deposited on the surface by the laser and $T_e$ and $T_l$ are the electronic and lattice temperatures, respectively. 

The lattice thermal conductivity is assumed to be negligible in comparison to the electron thermal conductivity and is usually neglected in TTM simulations \cite{Chen2007}. The electronic thermal heat capacity \textit{versus} temperature was taken from \cite{zhig2012} and fitted in order to have a clear relation with temperature over different temperature ranges.
\begin{figure}[H]
  \centering
  \begin{tabular}{@{}p{.495\linewidth}@{}p{.01\linewidth}@{}p{.495\linewidth}@{}}
	\subfigimg[width=\linewidth]{\textbf{\textcolor{red}{a)}}}{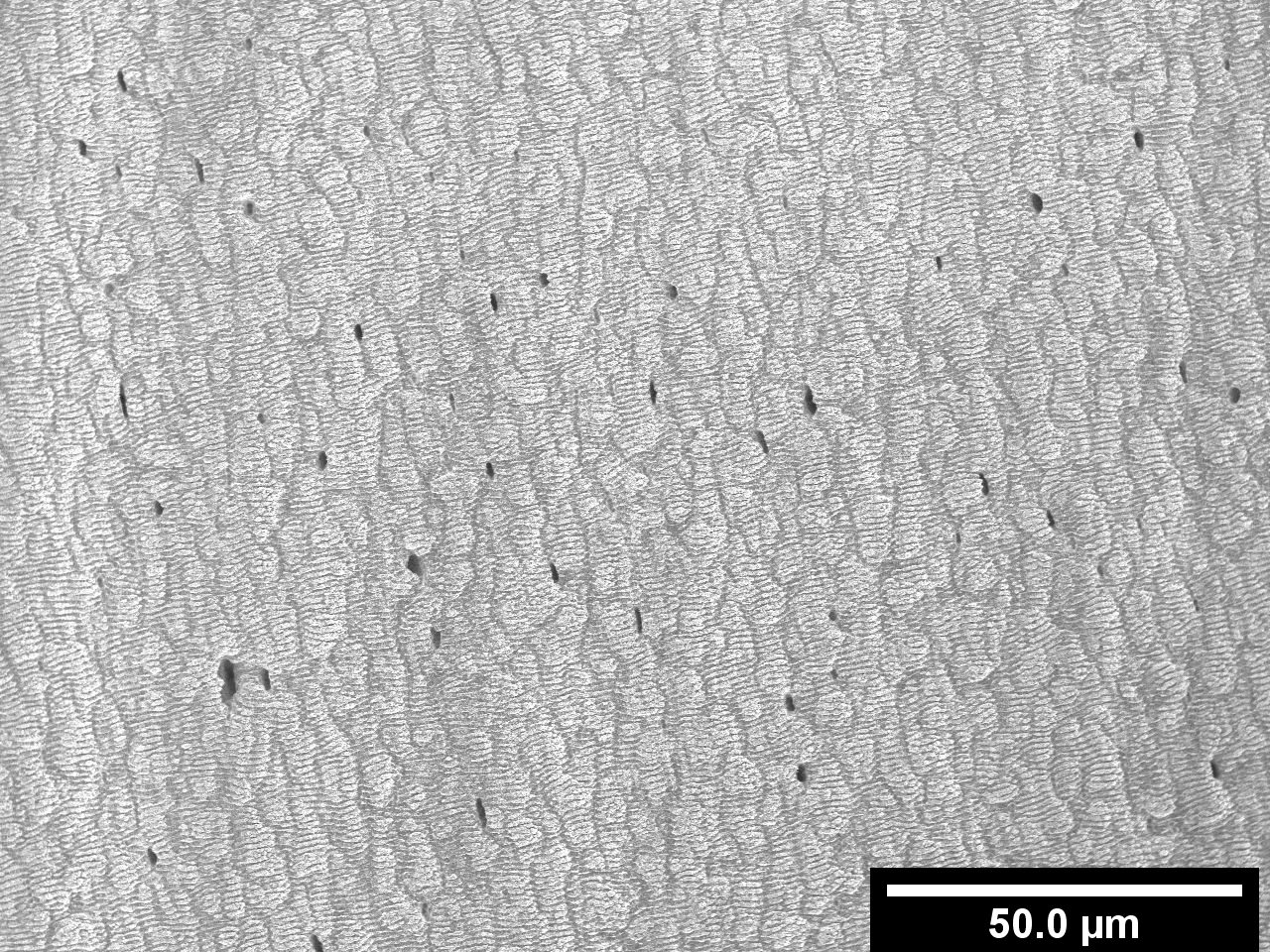}& &
		\subfigimg[width=\linewidth]{\textbf{\textcolor{red}{b)}}}{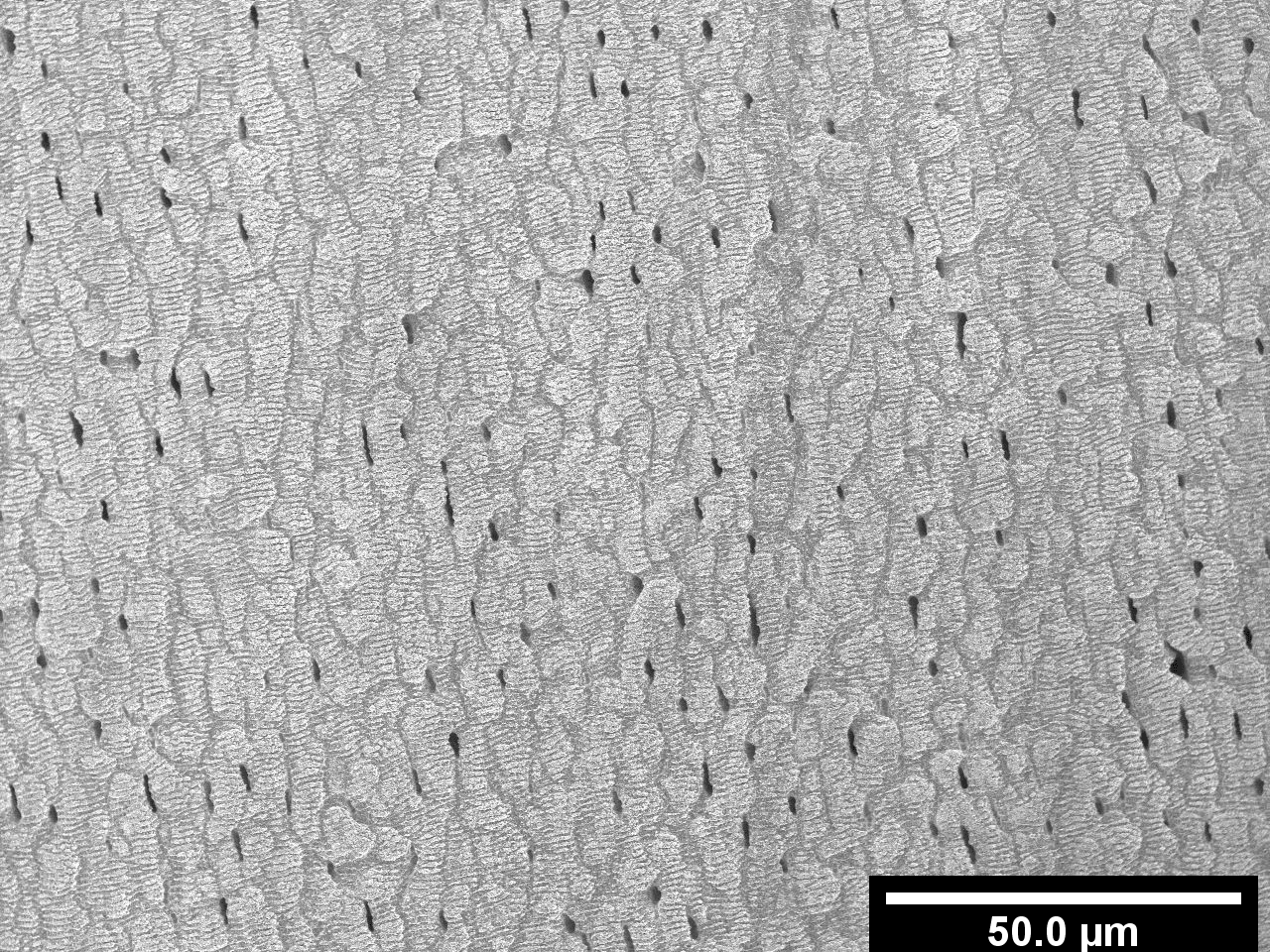}\\
			\subfigimg[width=\linewidth]{\textbf{\textcolor{red}{c)}}}{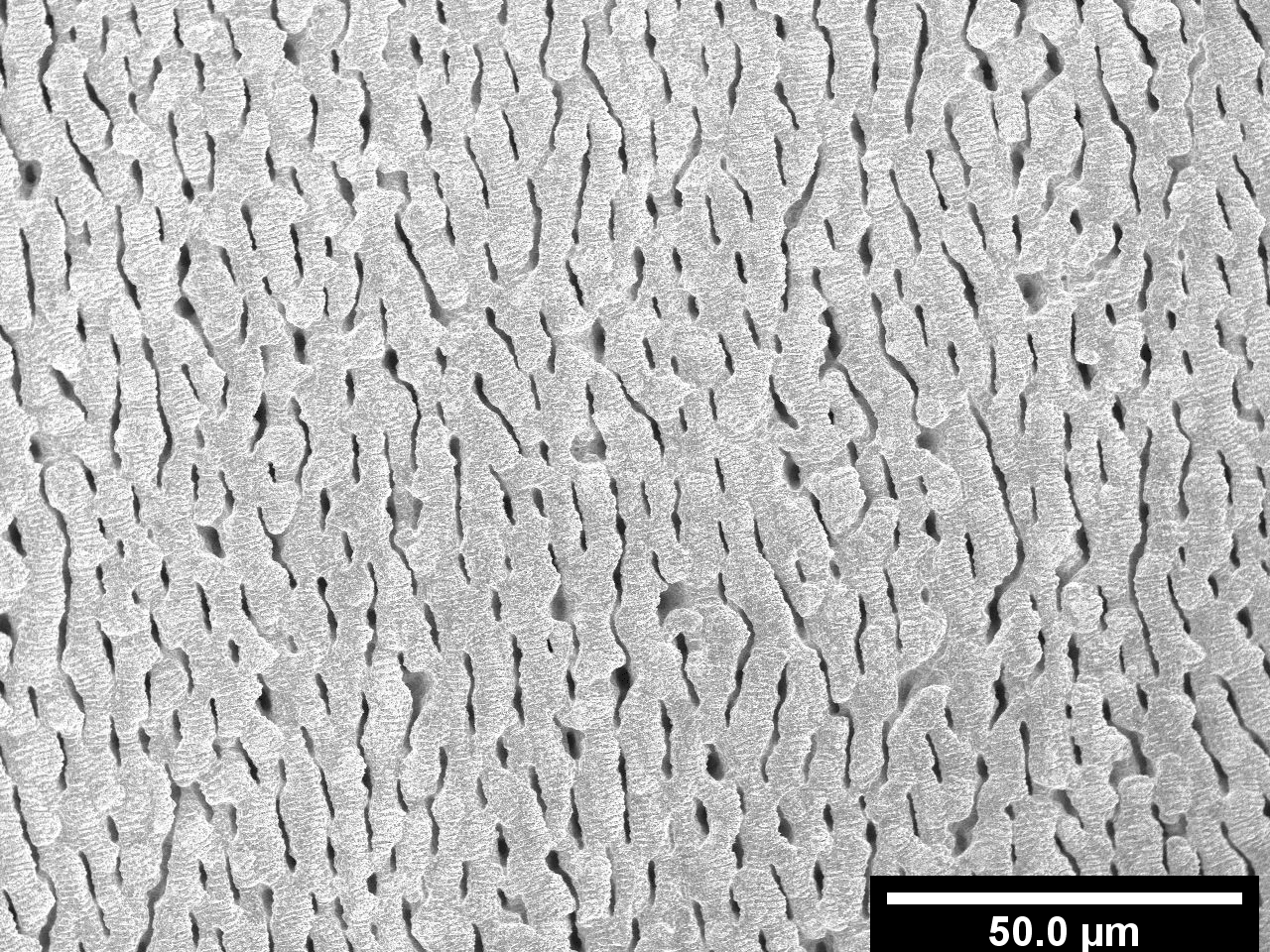}& &
				\subfigimg[width=\linewidth]{\textbf{\textcolor{red}{d)}}}{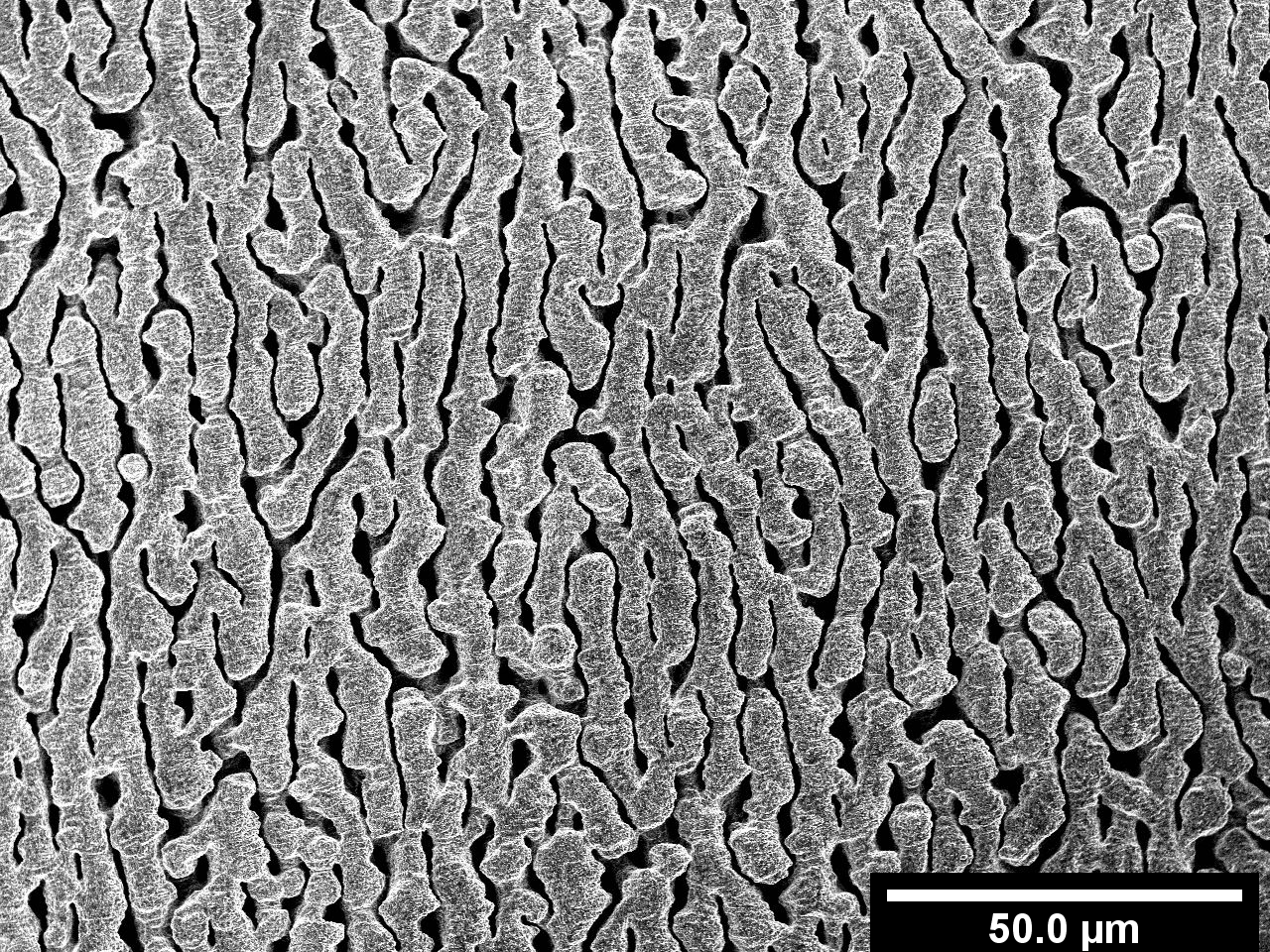}\\
					\subfigimg[width=\linewidth]{\textbf{\textcolor{red}{e)}}}{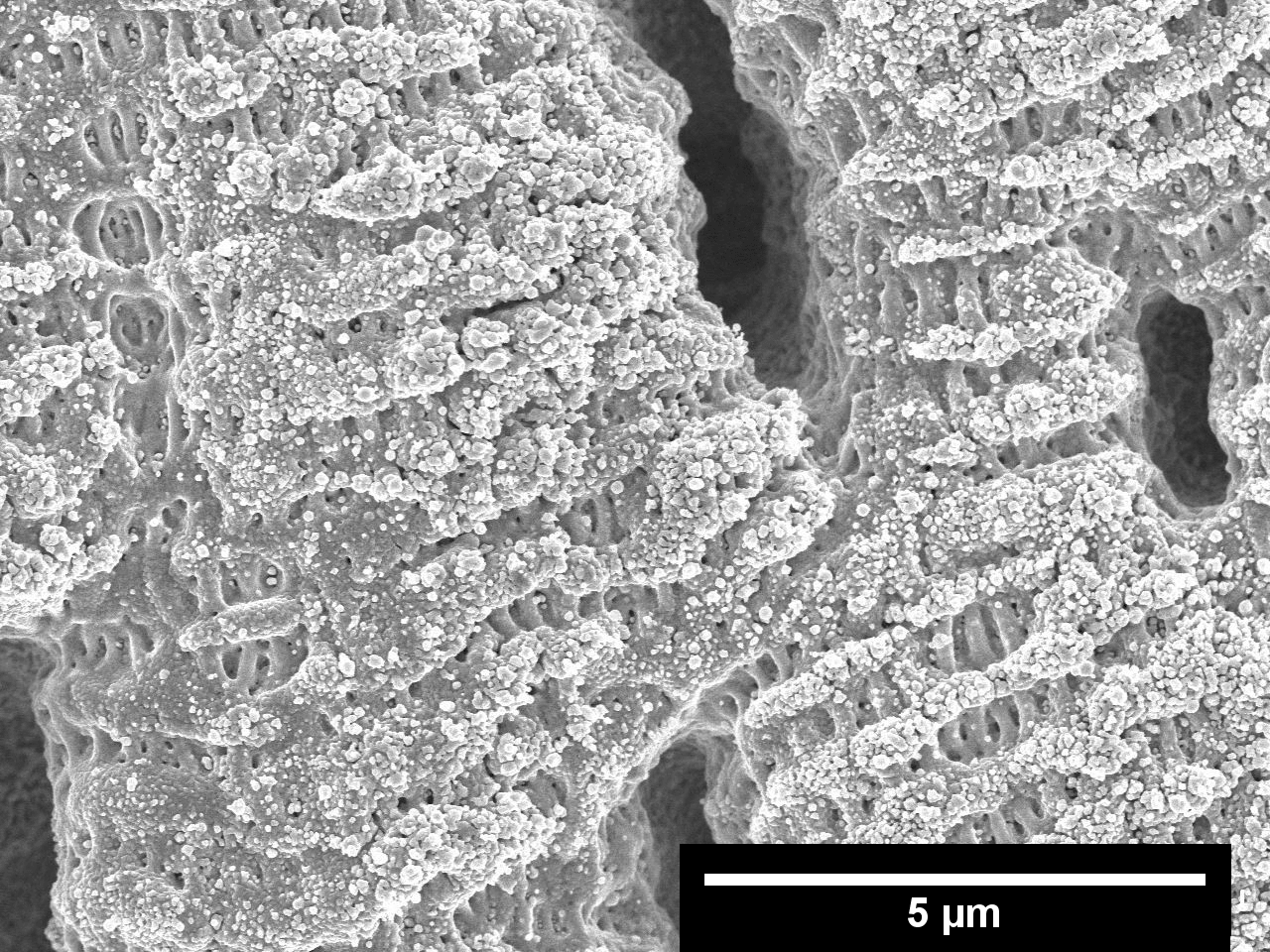}& &
			\subfigimg[width=\linewidth]{\textbf{\textcolor{red}{f)}}}{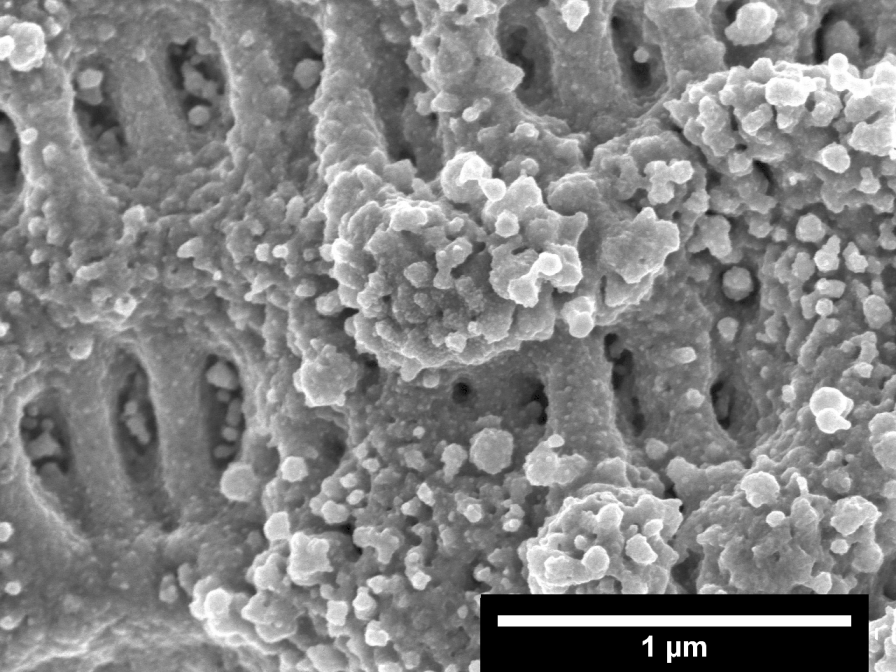}\\ 
				\subfigimg[width=\linewidth]{\textbf{\textcolor{red}{g)}}}{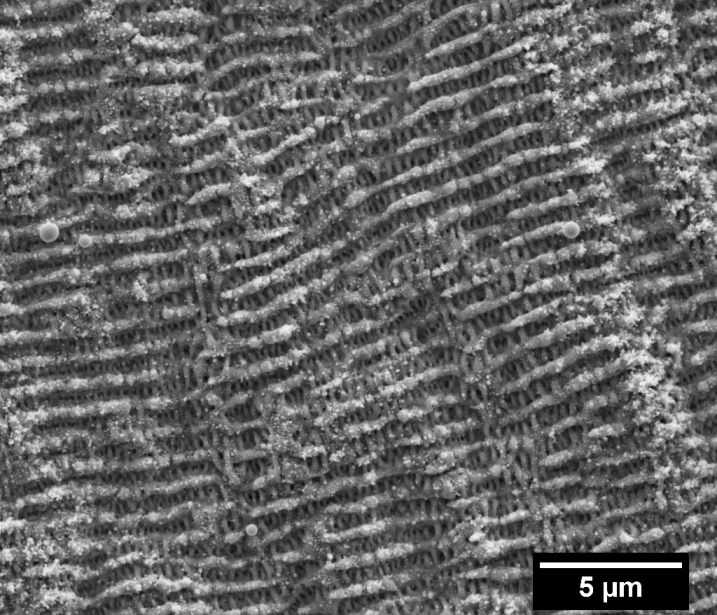}& &
					\subfigimg[width=\linewidth]{\textbf{\textcolor{red}{h)}}}{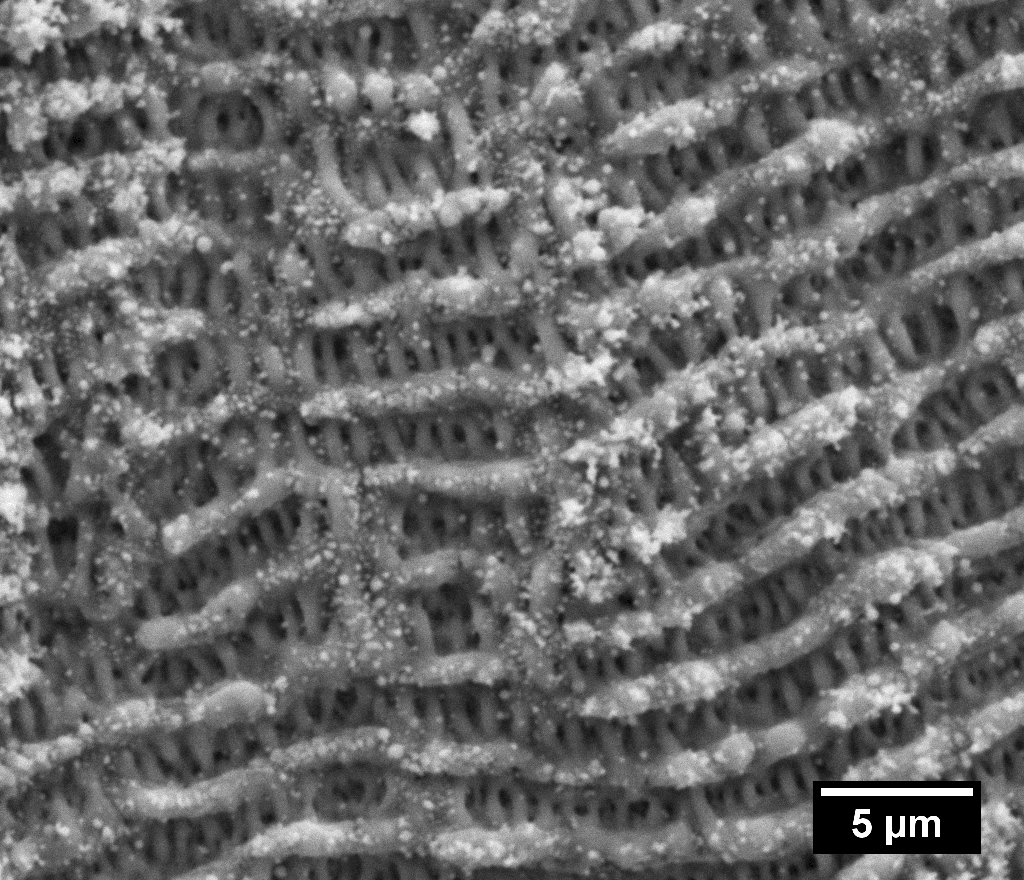}
  \end{tabular}
\caption{SEM images of colored surfaces following (a) 8, (b) 7, (c) 5 and (d) 4 pulse burst exposures with a writing speed of 20 mm/s and a laser fluence of 4.06 $J/cm^{2}$. Random melted structures are observed to form on the surface and coalesce into LLIPSS with a periodicity 10 times the wavelength and oriented parallel to the laser polarization. High magnification SEM images of colored surfaces using burst. (e,f) Surface processed using a laser fluence of 4.06 $J/cm^{2}$, 5 pulse bursts, marking speed of 20 mm/s and line spacing of 5 $\mu m$. (g,h) Surface processed using a laser fluence of 4.06 $J/cm^{2}$, line spacing of 5 $\mu m$, marking speed of 100 mm/s and 5 pulses burst. Well-defined HSFL are observed to form on the surface of silver in addition to traditional LIPSS (\textit{i.e.} LSFL).}
\label{fig4}
\end{figure}
\begin{figure}[H]
  \centering
  \begin{tabular}{@{}p{1\linewidth}@{}}
		\subfigimg[width=\linewidth]{\textbf{\textcolor{white}{}}}{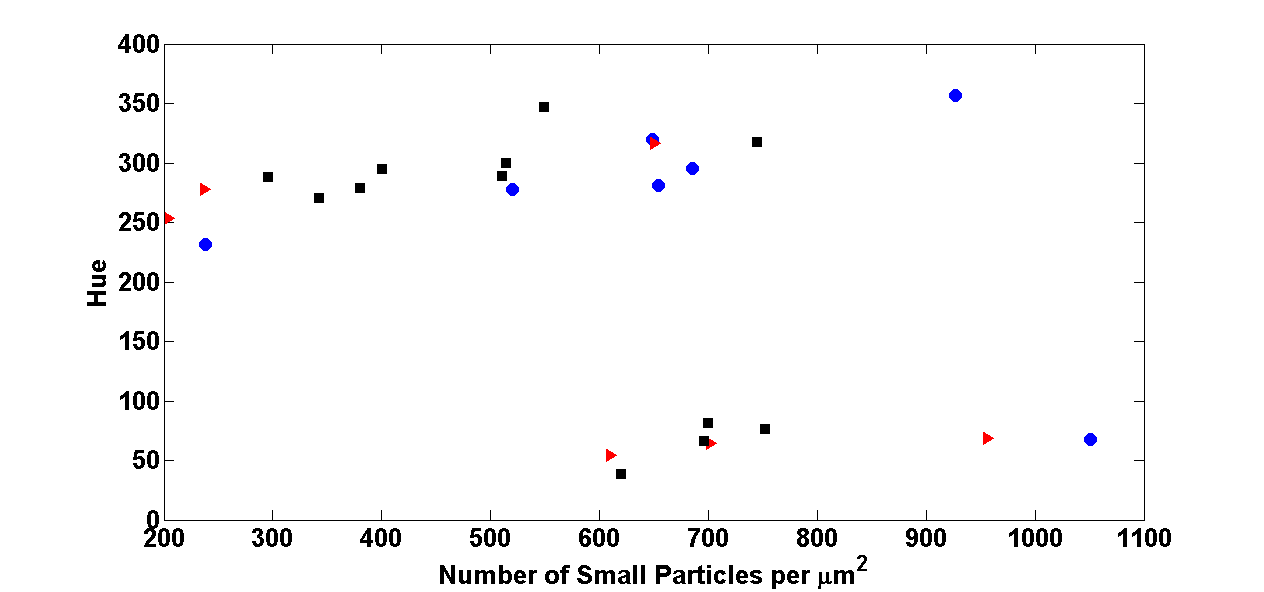}
	 \end{tabular}
\caption{Graph comparing the density of small particles on colored silver surfaces for the marking speeds of 11  ($\begingroup\color{blue}\bullet\endgroup$) and 50 mm/s (non-burst) ($\begingroup\color{red}\filledtriangleright\endgroup$) at 1.02 and 1.67 $J/cm^{2}$, respectively, and 20 mm/s with 5 pulse bursts at a laser fluence of 4.06 $J/cm^{2}$ ($\begingroup\color{black}\filledsquare\endgroup$) versus Hue (colors). The number of small NPs ($\sim$ 6.5 nm radius) is observed to be similar for each color.}
\label{fig5}
\end{figure}
The electronic thermal conductivity and electron-phonon coupling were calculated using the relations found in \cite{Chen2007}. The lattice thermal heat capacity was calculated in the high temperature limit applying the Debye model \cite{kittel}. The laser absorption by the material was calculated using the following relations \cite{Hu2009,Ren2012}:
\begin{equation}
S(z,t) = (1-R)\alpha I(t)e^{-\alpha z}
\end{equation}
and
\begin{equation}
I(t) = 0.94\frac{J_0}{\tau}e^{-2.77(\frac{t}{\tau})^2}
\end{equation}
where R is the surface reflectivity, $\alpha$ is the absorption coefficient, I is the laser intensity, $\tau$ is the pulse length, and $J_0$ is the laser fluence. The material reflectivity and absorption were continuously calculated using the temperature dependence of the electron relaxation time in the Drude + 2 Critical Point model \cite{Vial2008}. 

As the lattice temperature reached 0.9 times the critical temperature ($T_c$), critical point phase separation (phase explosion) occurs and material is removed from the surface \cite{Ren2013,Hu2009}. In our simulations, when material the reaches 0.9$T_c$ (in the lattice), the regions with this temperature are ignored for the remainder of the simulation and the layer below is assumed to be the new surface for the next laser pulse. 

The different colors created using a different number of pulses per burst for a fixed total energy is believed to come from the already hot surface that is encountered by subsequent laser pulses within a burst. Figure 6 (a), is a 1D TTM simulation of the temperature field with respect to time for a burst of 8 pulses, each separated by 12.8 ns. For simplicity each pulse within the burst was taken to be of equal amplitude affecting only the magnitude of the electron-phonon coupling. It can be clearly observed that the surface perceived by subsequent pulses is significantly hotter and increases with each pulse within the burst. In Figure 6 (b), we see during the laser irradiation that the electron temperature is significantly higher than that of the lattice. This is due to the assumption that the electrons absorb the laser energy before transferring it to the lattice. As energy diffuses into the bulk, the excited electrons transfer their energy to the lattice by collisions until the lattice accumulates more energy than the electron system near the surface. With time the system tends to equilibrium. In the case of burst, another pulse hits the surface before this equilibrium is reached.
\begin{figure}[H]
  \centering
  \begin{tabular}{@{}p{1\linewidth}@{}}
		\subfigimg[width=\linewidth]{\textbf{\textcolor{black}{(a)}}}{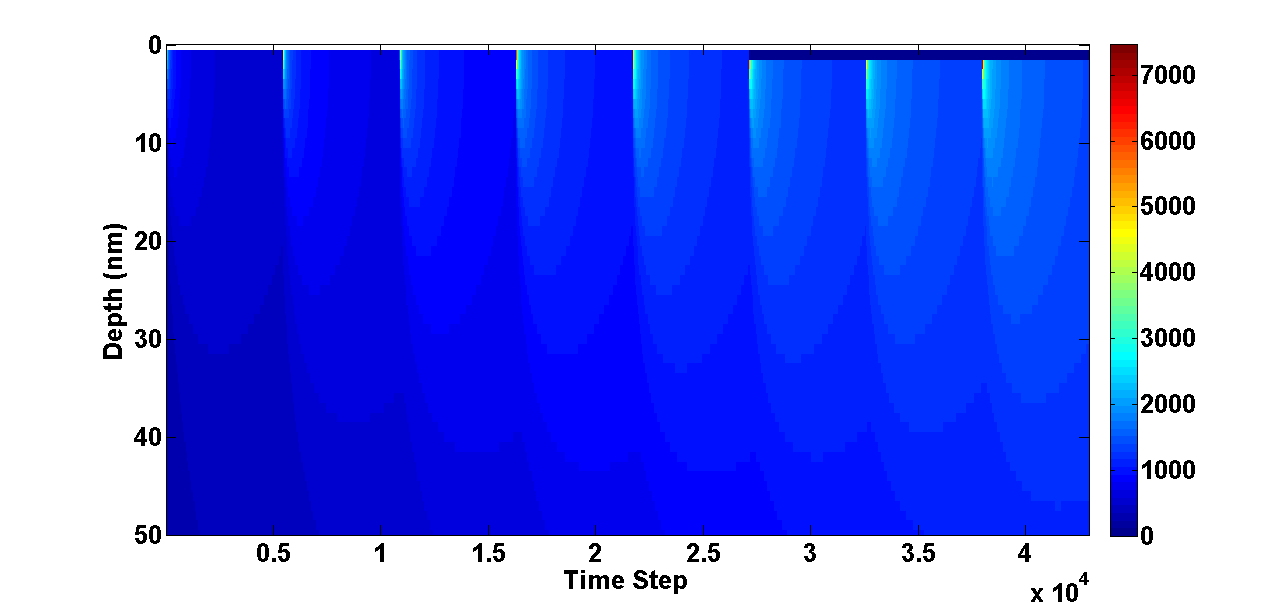}\\
				\subfigimg[width=\linewidth]{\textbf{\textcolor{black}{(b)}}}{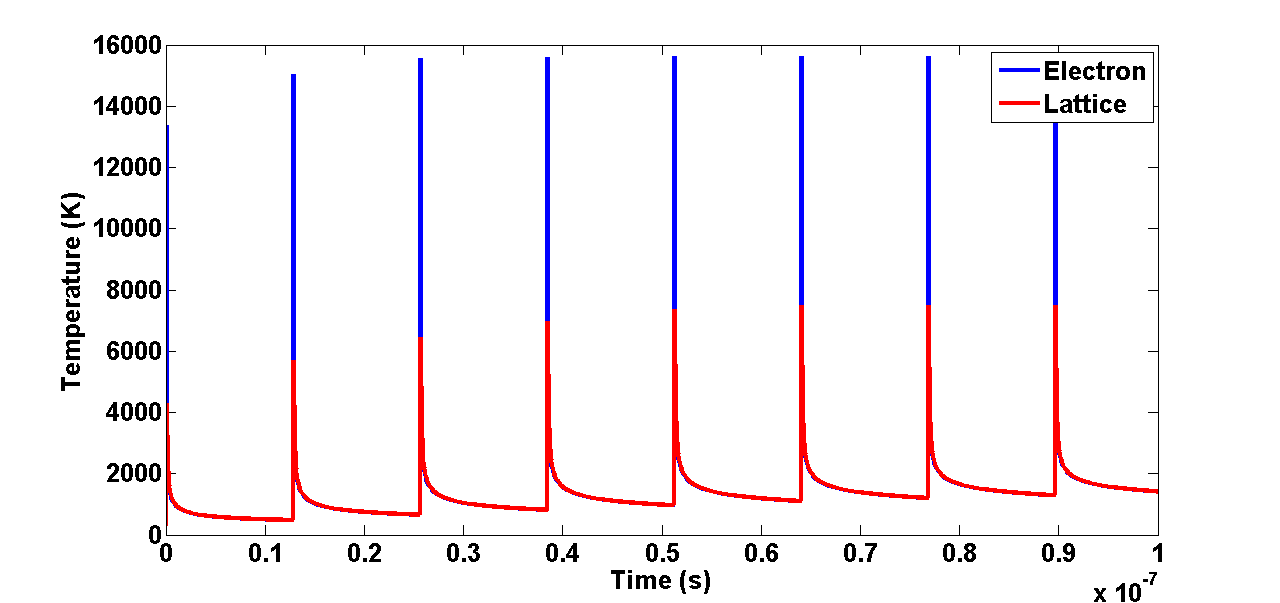}\\
				\subfigimg[width=\linewidth]{\textbf{\textcolor{black}{(c)}}}{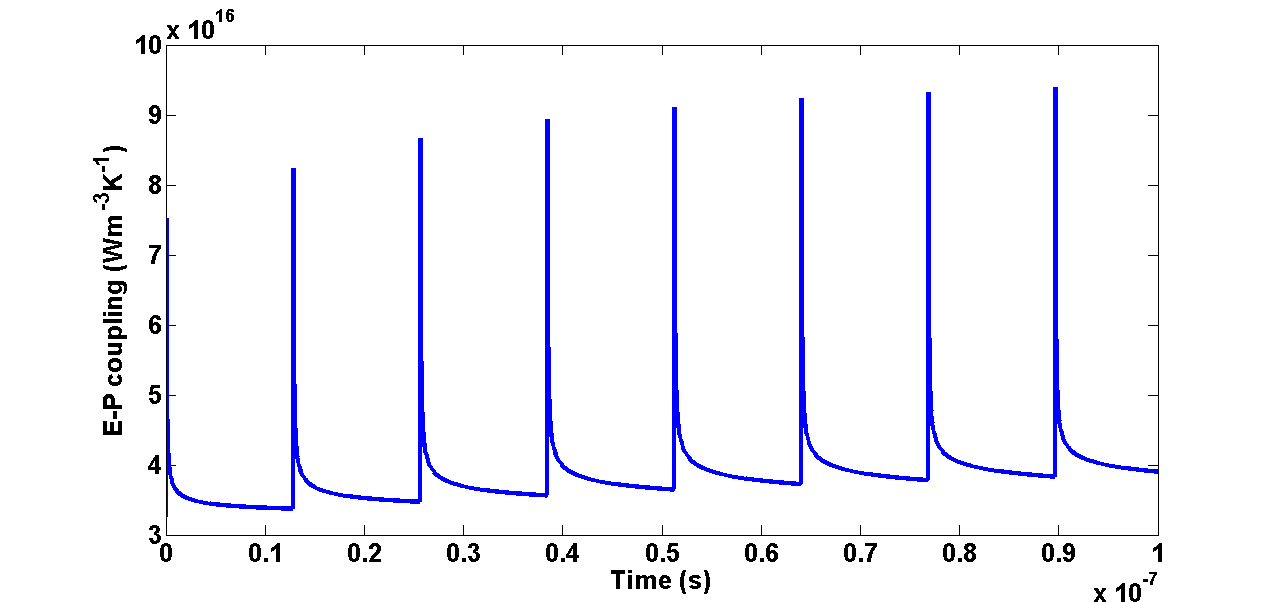}

	 \end{tabular}
\caption{(a) 1D temperature field with respect to time for an 8 pulse burst each separated by 12.8 ns with a total energy of 127 $\mu J$ simulated using TTM. (b,c) Graph showing electron and lattice temperature, and (c) electron-phonon coupling with respect to time.}
\label{fig6}
\end{figure}
The electron-phonon coupling values for burst in Fig. 6 (c) show an increase in electron-phonon coupling strength of up to $\sim$14$\%$ explaining the higher and better defined LIPSS observed in burst. The electron-phonon coupling coefficient is proportional to temperature. The pulse separation being much shorter than the thermal relaxation time of silver, heat accumulation is seen to occur in the metal causing the electron-phonon coupling to increase. The link between the LLIPSS and the higher electron-phonon coupling is however not clear. Other groups have observed similar structures from the accumulation of laser shots on a surface \cite{Ahmmed2014}, however, our experiments showed no such structures in the non-burst case even for the same local amount of shots indicating that the structures arise from interaction with the already melted surface. Higher magnification SEMs also reveal HSFL, Figure 4. Previous coloring of silver surfaces, without burst, fail in all instances to produce such structures on silver \cite{Guay2016a}. It is believed that the emergence of such structures with burst is directly linked to the increase of the electron-phonon coupling since it is known to play a vital role in the quality of ripple structures \cite{Wang2005,Vorobyev2013a}. The use of burst could prove to be highly effective in creating well defined LIPSS structures on metals that normally do not respond well to LIPSS formation.
	\begin{figure}[H]
  \centering
  \begin{tabular}{@{}p{1\linewidth}@{}}
		\subfigimg[width=\linewidth]{\textbf{\textcolor{black}{(a)}}}{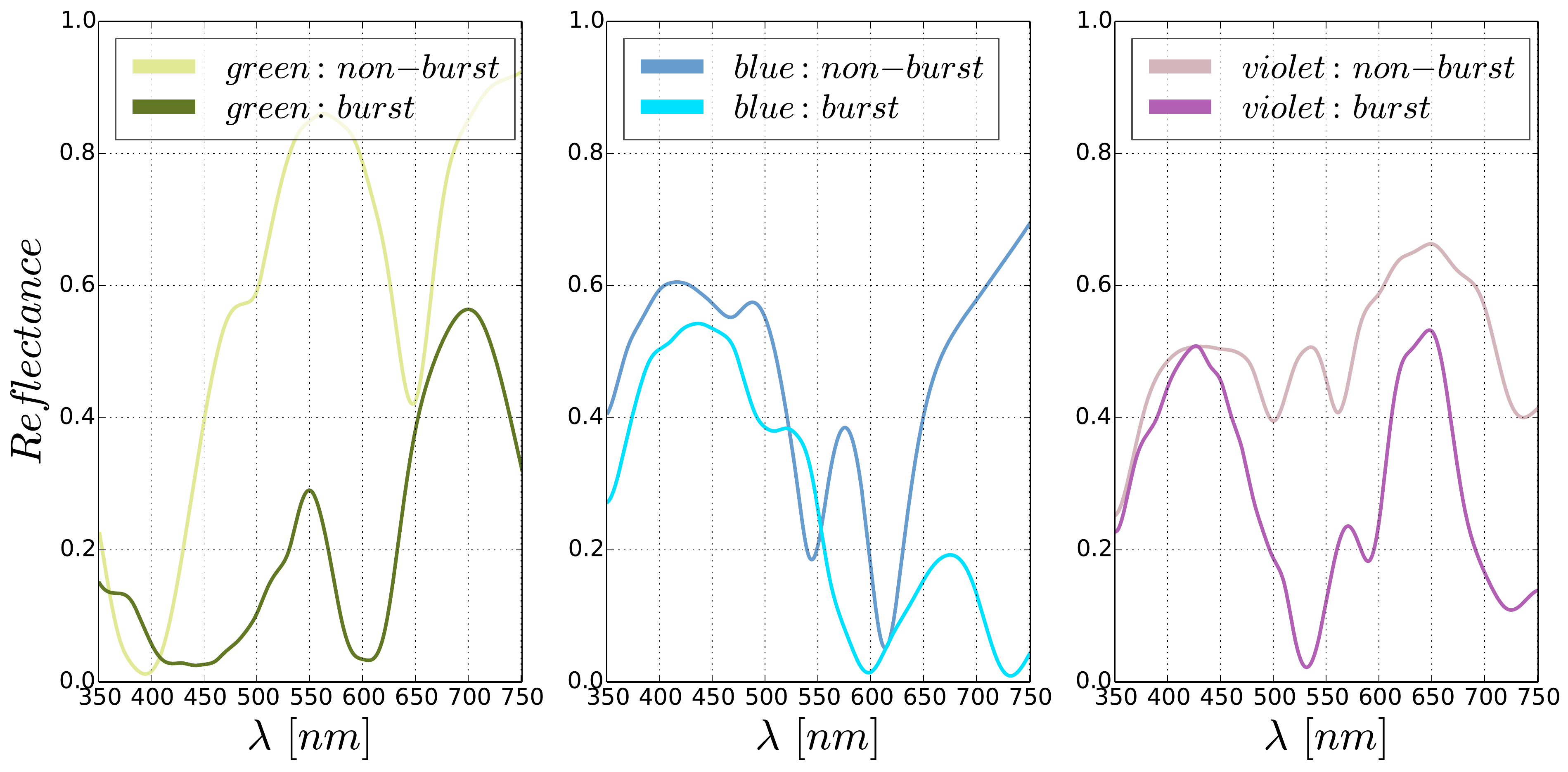}\\
		\subfigimg[width=\linewidth]{\textbf{\textcolor{black}{(b)}}}{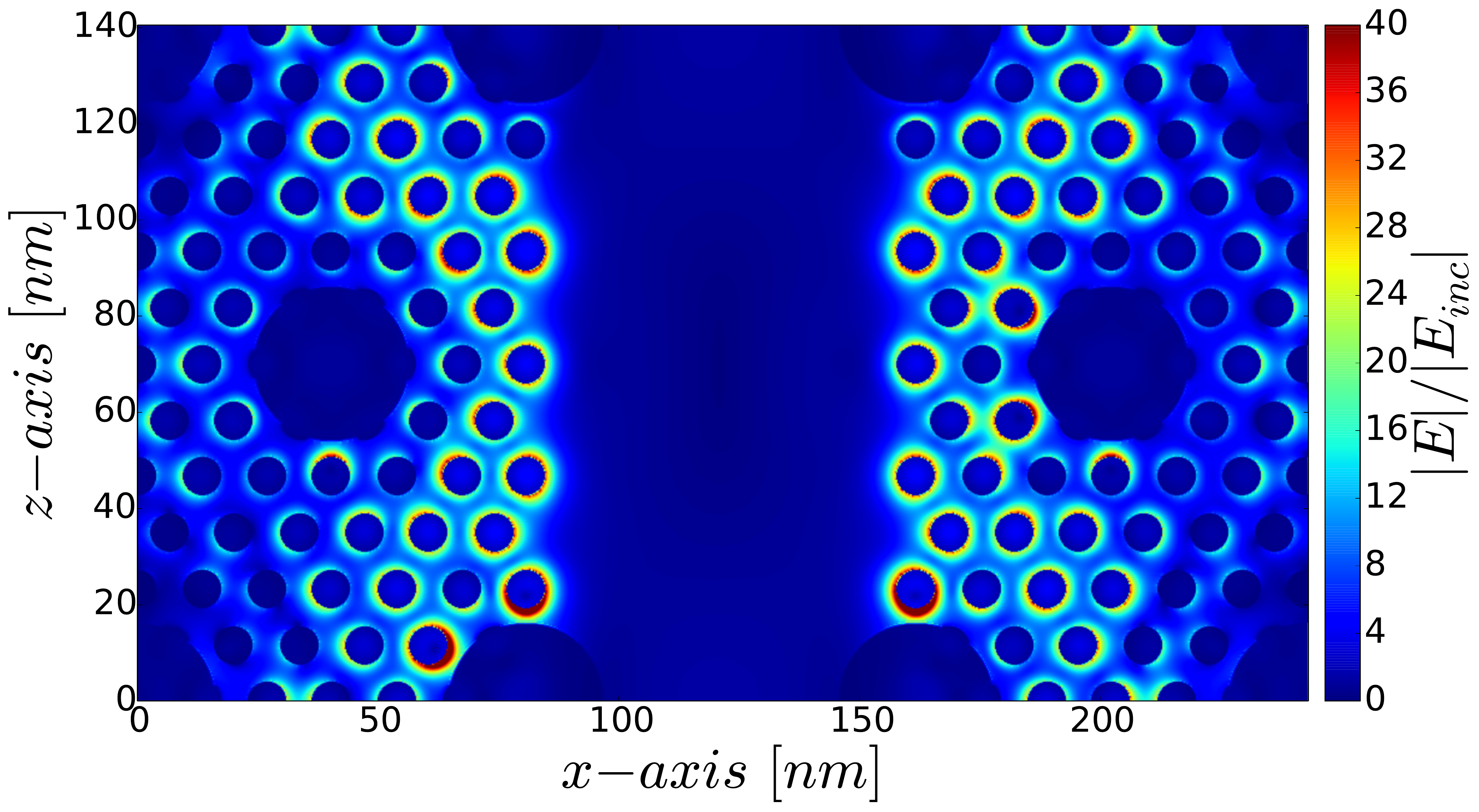}\\
		\subfigimg[width=\linewidth]{\textbf{\textcolor{black}{(c)}}}{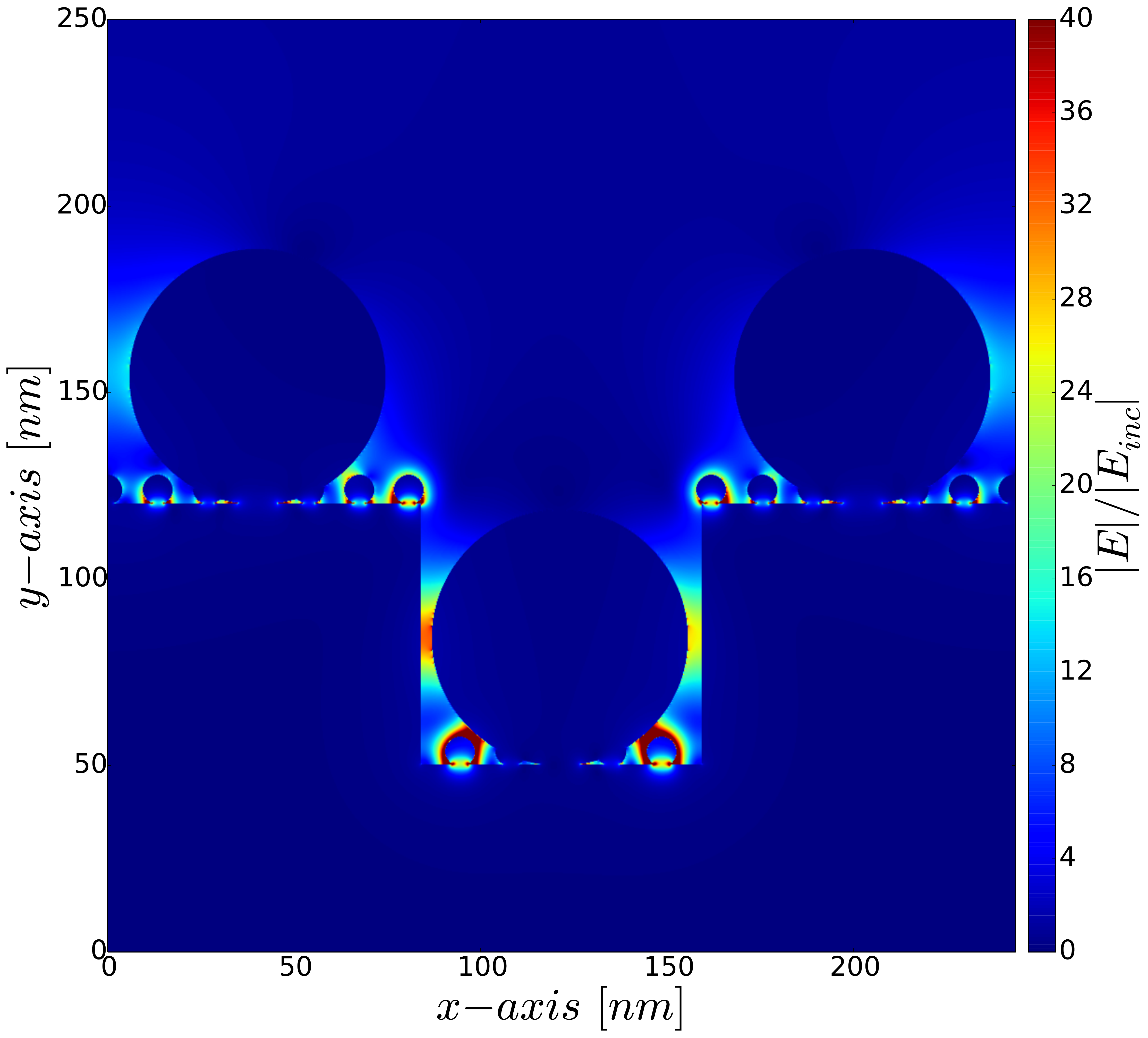}
	 \end{tabular}
\caption{(a) Graphs of reflectance versus wavelength computed using the FDTD method on burst and non-burst surfaces with the same particle density. FDTD simulations showing the electric field distribution on medium and small NPs sitting on HSFL structures at 730 nm in the (b) $xz$ and (c) $xy$ plane. The $xy$ plot is taken for 	$z$=0.5$\times z_{max}$ and $xz$ through the center of the small spheres on top of the burst structures.}
\label{fig7}
\end{figure}
\subsection{FDTD Simulations}
Following the approach in our previous work \cite{Guay2016a}, we utilized SEM statistical analyses of the nanoparticles distributed on the silver substrate for the FDTD simulations. The SEM analyses of the colored regions reveal the same nanoparticle density, for the same Hue value, following burst and non-burst processing of the surface, Figure 5. The only difference originate from the modulated silver surface. For LLIPSS, Chroma  (Lightness) values were observed to be always higher (lower) for an 8 pulse burst and, surface SEM images showed the LLIPSS to not yet be fully formed, Figure 4. From this observation the LLIPSS are believed to not play a significant role on the enhanced Chroma, but instead play a role on the overall Lightness, as the latter increases with a decreasing number of pulses per burst. In the case of LSFL, their dimensions are of the order of the laser wavelength which is much larger than the nanocluster size. For this reason, LSFL can be responsible for surface plasmon coupling, and angle-dependent coloring effects due to grating diffraction. However, we believe that they would not affect significantly the resonance condition of the heterogeneous nanoclusters of small and medium size nanoparticles (NPs), which contribute angle-independent color formation by selective absorption. In addition, from the manipulation of the coin and its observation under various viewing angles, the colors were found to be angle-independent suggesting that selective absorption dominates over light diffraction. Thus, only HSFL are considered in the 3D-FDTD simulations as their dimensions introduce resonance conditions in the optical range, and would therefore participate in the color formation process, in agreement with groups that showed angle-independent coloring of metals using periodic structures below the diffraction of visible light.
\\
\\
Similar to our previous work, we chose the inter-particle distance of the medium NPs as an integer multiple of the inter-particle distance of the small NPs, in order to apply periodic boundary conditions (PBCs) \cite{Guay2016a}. To reproduce the surface following burst processing, we introduce a modulation of the surface along the $x$-direction with crevices of width $\sim$100 nm (based on SEM) and height $\sim$50 nm to simulate the presence of the HSFL. PBCs are applied in the $x$ and $z$ directions. For a flat surface (\textit{non-burst}) the reflectance is polarization independent due to the distribution of NPs on a hexagonal lattice; this does not hold in the burst case because of the non-flat surface. In Fig. 7(a), we show the reflectance for an x-polarized excitation to include the effect of the crevices (\textit{i.e.} HSFL). The colors perceived would nonetheless remain relatively angle-independent as the color observed would be an average of all the different polarization component incident on the structured surface.

In Fig. 7(a), we show the reflectance associated with three colors obtained for non-burst simulations, \textit{e.g}, green, blue and magenta, and the reflectance resulting from the same NP density distributed on the surface modulated by HSFL crevices. This leads to an increase in Chroma from 31.8 to 40.6 or $\sim$27$\%$ for green, from 41.6 to 55 or $\sim$32$\%$ for blue, and from 19.3 to 64.4 or $\sim$233$\%$ for magenta. The color was rendered using Matlab code to convert the reflectance to color, and the RGB values are used to draw the corresponding lines on Fig. 7 (a). The presence of the crevices in the burst case increase the Chroma values for the three different colors with respect to the non-burst case (\textit{i.e.} flat surface), as observed in experiments. We observed experimentally, overall, a maximum increase of $\sim$68$\%$ for green (90-150 Hue), $\sim$61$\%$ for blue (210-270 Hue) and $\sim$22$\%$ for magenta (270-330 Hue).
\\
\\
The sub-wavelength HSFL influence the color formation process because of the enhanced absorption in the crevices. The plasmon resonances are enriched by the presence of the $x$-periodicity, which introduces new surface collective resonances at higher wavelengths. This is revealed by the absorption dips at higher wavelengths in Figure 7 (a), and confirmed by the field distribution at 730 nm as seen in the blue color case in Figure 7 (b,c). The latter figure shows the absolute value of the electric field on an $xz$ plane cut through the middle of the small NPs; in the central part of the figure there are no NPs since they are located at a lower height inside the crevice. 

\section{Conclusion}
Colors covering the full spectrum were obtained using closely time-spaced laser pulses (burst mode). The use of burst mode to color silver is observed increase Chroma values by $\sim$50$\%$ over most of the color spectrum and broadens the lightness range by $\sim$60$\%$ compared to non-burst mode. These increases are accompanied by the creation of 3 kinds different laser-induced periodic surface structures on the surface of silver. One of these structures is 10 times the wavelength of light and cannot be explained by conventional theories. The LLIPSS is observed to increase the Lightness of the colors. Two-Temperature model simulations show a significant increase in the electron-phonon coupling using burst and is believed to the responsible for the well defined LIPSS and HFLS on the surface of silver following picosecond laser exposure. FDTD modeling of the representative surfaces corroborate the involvement of surface plasmons in the color formation process and show the importance of the crevices (\textit{i.e. HSFL}) to enhanced absorption and enriched plasmon resonances. The presence of the LSFL on the surface and the absence of angle-dependent colors suggest that selective absorption dominates over diffraction.

\appendix
\subsection*{Acknowledgements}
We acknowledge the Royal Canadian Mint, the Natural Sciences and Engineering Council of Canada, the Canada Research Chairs program, the Southern Ontario Smart Computing Innovation Platform (SOSCIP), and SciNet. We would like to acknowledge COOP students Graham Jerrey Rivers Killaire and Mael Chow-Cloutier at the University of Ottawa. 

\subsection*{Competing financial interests}
The authors declare that they have no competing financial interests.
%\subsection*{Correspondence}
%Correspondence and requests for materials
%should be addressed to Jean-Michel Guay~(email: guay.jeanmichel@gmail.com).

\bibliography{references}{}

\begin{thebibliography}{10}
\expandafter\ifx\csname url\endcsname\relax
  \def\url#1{\texttt{#1}}\fi
\expandafter\ifx\csname urlprefix\endcsname\relax\def\urlprefix{URL }\fi
\providecommand{\bibinfo}[2]{#2}
\providecommand{\eprint}[2][]{\url{#2}}

\bibitem{Vorobyev2013a}
\bibinfo{author}{Vorobyev, A.~Y.} \& \bibinfo{author}{Guo, C.}
\newblock \bibinfo{title}{{Direct femtosecond laser surface
  nano/microstructuring and its applications}}.
\newblock \emph{\bibinfo{journal}{Laser \& Photonics Reviews}}
  \textbf{\bibinfo{volume}{7}}, \bibinfo{pages}{385--407}
  (\bibinfo{year}{2013}).

\bibitem{Fan2014}
\bibinfo{author}{Fan, P.} \emph{et~al.}
\newblock \bibinfo{title}{{Angle-independent colorization of copper surfaces by
  simultaneous generation of picosecond-laser-induced nanostructures and
  redeposited nanoparticles}}.
\newblock \emph{\bibinfo{journal}{Journal of Applied Physics}}
  \textbf{\bibinfo{volume}{115}}, \bibinfo{pages}{124302}
  (\bibinfo{year}{2014}).

\bibitem{Guay2016a}
\bibinfo{author}{Guay, J.-M.} \emph{et~al.}
\newblock \bibinfo{title}{{Topographical coloured plasmonic coins}}
  (\bibinfo{year}{2016}).
\newblock \eprint{1609.02874}.

\bibitem{Salvador2012}
\bibinfo{author}{Salvador, M.} \emph{et~al.}
\newblock \bibinfo{title}{{Electron Accumulation on Metal Nanoparticles in
  Plasmon-Enhanced Organic Solar Cells}}.
\newblock \emph{\bibinfo{journal}{ACS nano}} \textbf{\bibinfo{volume}{6}},
  \bibinfo{pages}{10024--10032} (\bibinfo{year}{2012}).

\bibitem{Rai2015}
\bibinfo{author}{Rai, M.}, \bibinfo{author}{Ingle, A.~P.},
  \bibinfo{author}{Birla, S.}, \bibinfo{author}{Yadav, A.} \&
  \bibinfo{author}{Santos, C. A.~D.}
\newblock \bibinfo{title}{{Strategic role of selected noble metal nanoparticles
  in medicine.}}
\newblock \emph{\bibinfo{journal}{Critical reviews in microbiology}}
  \bibinfo{pages}{1--24} (\bibinfo{year}{2015}).

\bibitem{Huang2011}
\bibinfo{author}{Huang, X.} \& \bibinfo{author}{El-Sayed, M.~a.}
\newblock \bibinfo{title}{{Plasmonic photo-thermal therapy (PPTT)}}.
\newblock \emph{\bibinfo{journal}{Alexandria Journal of Medicine}}
  \textbf{\bibinfo{volume}{47}}, \bibinfo{pages}{1--9} (\bibinfo{year}{2011}).

\bibitem{Catchpole2008}
\bibinfo{author}{Catchpole, K.~R.} \& \bibinfo{author}{Polman, A.}
\newblock \bibinfo{title}{{Plasmonic solar cells}}.
\newblock \emph{\bibinfo{journal}{Optics express}}
  \textbf{\bibinfo{volume}{16}}, \bibinfo{pages}{21793--21800}
  (\bibinfo{year}{2008}).

\bibitem{Mie}
\bibinfo{author}{{G. Mie}}.
\newblock \bibinfo{title}{{Beitr\"{a}ge zur Optik tr\"{u}ber Medien, speziell
  kolloidaler Metall\"{o}sungen}}.
\newblock \emph{\bibinfo{journal}{Annalen Der Physik}}
  \bibinfo{pages}{377--445} (\bibinfo{year}{1908}).

\bibitem{Doyle1989}
\bibinfo{author}{Doyle, W.}
\newblock \bibinfo{title}{{Optical properties of a suspension of metal
  spheres.}}
\newblock \emph{\bibinfo{journal}{Physical review. B, Condensed matter}}
  \textbf{\bibinfo{volume}{39}}, \bibinfo{pages}{9852--9858}
  (\bibinfo{year}{1989}).

\bibitem{Murray2007}
\bibinfo{author}{Murray, W.} \& \bibinfo{author}{Barnes, W.}
\newblock \bibinfo{title}{{Plasmonic Materials}}.
\newblock \emph{\bibinfo{journal}{Advanced Materials}}
  \textbf{\bibinfo{volume}{19}}, \bibinfo{pages}{3771--3782}
  (\bibinfo{year}{2007}).

\bibitem{Gallinet2015}
\bibinfo{author}{Duempelmann, L.}, \bibinfo{author}{Casari, D.},
  \bibinfo{author}{Luu-Dinh, A.}, \bibinfo{author}{Gallinet, B.} \&
  \bibinfo{author}{Novotny, L.}
\newblock \bibinfo{title}{{Color Rendering Plasmonic Aluminum Substrates with
  Angular Symmetry}}.
\newblock \emph{\bibinfo{journal}{ACS nano}}  (\bibinfo{year}{2015}).

\bibitem{Roberts2014}
\bibinfo{author}{Roberts, A.~S.}, \bibinfo{author}{Pors, A.},
  \bibinfo{author}{Albrektsen, O.} \& \bibinfo{author}{Bozhevolnyi, S.~I.}
\newblock \bibinfo{title}{{Subwavelength Plasmonic Color Printing Protected for
  Ambient Use}}.
\newblock \emph{\bibinfo{journal}{Nano Letters}} \textbf{\bibinfo{volume}{14}},
  \bibinfo{pages}{783--787} (\bibinfo{year}{2014}).

\bibitem{Tan2014}
\bibinfo{author}{Tan, S.~J.} \emph{et~al.}
\newblock \bibinfo{title}{{Plasmonic Color Palettes for Photorealistic Printing
  with Aluminum Nanostructures}}.
\newblock \emph{\bibinfo{journal}{Nano Letters}} \textbf{\bibinfo{volume}{14}},
  \bibinfo{pages}{4023--4029} (\bibinfo{year}{2014}).

\bibitem{Cheng2015}
\bibinfo{author}{Cheng, F.} \emph{et~al.}
\newblock \bibinfo{title}{{Aluminum plasmonic metamaterials for structural
  color printing}}.
\newblock \emph{\bibinfo{journal}{Optics express}}
  \textbf{\bibinfo{volume}{23}}, \bibinfo{pages}{23279--23285}
  (\bibinfo{year}{2015}).

\bibitem{Clausen2014}
\bibinfo{author}{Clausen, J.~S.} \emph{et~al.}
\newblock \bibinfo{title}{{Plasmonic Metasurfaces for Coloration of Plastic
  Consumer Products}}.
\newblock \emph{\bibinfo{journal}{Nano Letters}} \textbf{\bibinfo{volume}{14}},
  \bibinfo{pages}{4499--4504} (\bibinfo{year}{2014}).

\bibitem{Huang2009}
\bibinfo{author}{Huang, M.}, \bibinfo{author}{Zhao, F.},
  \bibinfo{author}{Cheng, Y.}, \bibinfo{author}{Xu, N.} \& \bibinfo{author}{Xu,
  Z.}
\newblock \bibinfo{title}{{Origin of laser-induced near-subwavelength ripples:
  interference between surface plasmons and incident laser.}}
\newblock \emph{\bibinfo{journal}{ACS nano}} \textbf{\bibinfo{volume}{3}},
  \bibinfo{pages}{4062--70} (\bibinfo{year}{2009}).

\bibitem{Fauchet1982}
\bibinfo{author}{Fauchet, P.~M.}
\newblock \bibinfo{title}{{Surface ripples on silicon and gallium arsenide
  under picosecond laser illumination}}.
\newblock \emph{\bibinfo{journal}{Applied Physics Letters}}
  \textbf{\bibinfo{volume}{40}}, \bibinfo{pages}{824} (\bibinfo{year}{1982}).

\bibitem{Sipe1983}
\bibinfo{author}{Sipe, J.~E.}, \bibinfo{author}{Young, J.~F.} \&
  \bibinfo{author}{Preston, J.~S.}
\newblock \bibinfo{title}{{Laser-induced periodic surface structure. I.
  Theory}}.
\newblock \emph{\bibinfo{journal}{Physical Review B}}
  \textbf{\bibinfo{volume}{27}} (\bibinfo{year}{1983}).

\bibitem{phenon}
\bibinfo{author}{van Driel, H.}, \bibinfo{author}{Sipe, J.} \&
  \bibinfo{author}{Young, J.~F.}
\newblock \bibinfo{title}{{Laser-Induced Periodic Surface on Solids: A
  Universal Phenomenon}}.
\newblock \emph{\bibinfo{journal}{Physical Review Letters}}
  \textbf{\bibinfo{volume}{49}}, \bibinfo{pages}{1955--1960}
  (\bibinfo{year}{1982}).

\bibitem{Yao2012a}
\bibinfo{author}{Yao, J.-W.} \emph{et~al.}
\newblock \bibinfo{title}{{High spatial frequency periodic structures induced
  on metal surface by femtosecond laser pulses.}}
\newblock \emph{\bibinfo{journal}{Optics express}}
  \textbf{\bibinfo{volume}{20}}, \bibinfo{pages}{905--11}
  (\bibinfo{year}{2012}).

\bibitem{Bonse2011}
\bibinfo{author}{Bonse, J.}, \bibinfo{author}{Rosenfeld, a.} \&
  \bibinfo{author}{Kr\"{u}ger, J.}
\newblock \bibinfo{title}{{Implications of transient changes of optical and
  surface properties of solids during femtosecond laser pulse irradiation to
  the formation of laser-induced periodic surface structures}}.
\newblock \emph{\bibinfo{journal}{Applied Surface Science}}
  \textbf{\bibinfo{volume}{257}}, \bibinfo{pages}{5420--5423}
  (\bibinfo{year}{2011}).

\bibitem{Hartmann2007}
\bibinfo{author}{Hartmann, C.} \emph{et~al.}
\newblock \bibinfo{title}{{Investigation on laser micro ablation of metals
  using ns-multi-pulses}}.
\newblock \emph{\bibinfo{journal}{Journal of Physics: Conference Series}}
  \textbf{\bibinfo{volume}{59}}, \bibinfo{pages}{440--444}
  (\bibinfo{year}{2007}).

\bibitem{Herman1999}
\bibinfo{author}{Herman, P.~R.}, \bibinfo{author}{Oettl, A.},
  \bibinfo{author}{Chen, K.~P.} \& \bibinfo{author}{Marjoribanks, R.~S.}
\newblock \bibinfo{title}{{Laser micromachining of transparent fused silica
  with 1-ps pulses and pulse trains}}.
\newblock \emph{\bibinfo{journal}{SPIE}} \textbf{\bibinfo{volume}{3616}},
  \bibinfo{pages}{148--155} (\bibinfo{year}{1999}).

\bibitem{Ren2013}
\bibinfo{author}{Ren, Y.}, \bibinfo{author}{Cheng, C.}, \bibinfo{author}{Chen,
  J.}, \bibinfo{author}{Zhang, Y.} \& \bibinfo{author}{Tzou, D.}
\newblock \bibinfo{title}{{Thermal ablation of metal films by femtosecond laser
  bursts}}.
\newblock \emph{\bibinfo{journal}{International Journal of Thermal Sciences}}
  \textbf{\bibinfo{volume}{70}}, \bibinfo{pages}{32--40}
  (\bibinfo{year}{2013}).

\bibitem{Hu2009}
\bibinfo{author}{Hu, W.}, \bibinfo{author}{Shin, Y.~C.} \&
  \bibinfo{author}{King, G.}
\newblock \bibinfo{title}{{Modeling of multi-burst mode pico-second laser
  ablation for improved material removal rate}}.
\newblock \emph{\bibinfo{journal}{Applied Physics A}}
  \textbf{\bibinfo{volume}{98}}, \bibinfo{pages}{407--415}
  (\bibinfo{year}{2009}).

\bibitem{Wang2005}
\bibinfo{author}{Wang, J.} \& \bibinfo{author}{Guo, C.}
\newblock \bibinfo{title}{{Ultrafast dynamics of femtosecond laser-induced
  periodic surface pattern formation on metals}}.
\newblock \emph{\bibinfo{journal}{Applied Physics Letters}}
  \textbf{\bibinfo{volume}{87}}, \bibinfo{pages}{251914}
  (\bibinfo{year}{2005}).

\bibitem{Jandeleit1996}
\bibinfo{author}{Jandeleit, J.}, \bibinfo{author}{Urbasch, G.},
  \bibinfo{author}{Hoffmann, H.~D.}, \bibinfo{author}{Treusch, H.-G.} \&
  \bibinfo{author}{Kreutz, E.~W.}
\newblock \bibinfo{title}{{Picosecond laser ablation of thin copper films}}.
\newblock \emph{\bibinfo{journal}{Applied Physics A: Materials Science \&
  Processing}} \textbf{\bibinfo{volume}{63}}, \bibinfo{pages}{117--121}
  (\bibinfo{year}{1996}).

\bibitem{Kirkwood2007}
\bibinfo{author}{Kirkwood, S.}
\newblock \emph{\bibinfo{title}{{Characterization of metal and semiconductor
  nanomilling at near threshold intensities using femtosecond laser pulses}}}.
\newblock Ph.D. thesis, \bibinfo{school}{Alberta} (\bibinfo{year}{2007}).

\bibitem{Taflove2005}
\bibinfo{author}{Taflove, A.} \& \bibinfo{author}{Hagness, S.}
\newblock \emph{\bibinfo{title}{{Computational Electrodynamics: The
  Finite-Difference Time-Domain Method}}} (\bibinfo{publisher}{Artech House},
  \bibinfo{year}{2005}), \bibinfo{edition}{3} edn.

\bibitem{Taflove2013}
\bibinfo{author}{Taflove, A.}, \bibinfo{author}{Oskooi, A.} \&
  \bibinfo{author}{Johnson, S.~G.}
\newblock \emph{\bibinfo{title}{{Advances in FDTD Computational
  Electrodynamics: Photonics and Nanotechnology}}} (\bibinfo{publisher}{Artech
  House}, \bibinfo{year}{2013}).

\bibitem{Lesina2015}
\bibinfo{author}{Cal\`{a}~Lesina, A.}, \bibinfo{author}{Vaccari, A.},
  \bibinfo{author}{Berini, P.} \& \bibinfo{author}{Ramunno, L.}
\newblock \bibinfo{title}{{On the convergence and accuracy of the FDTD method
  for nanoplasmonics}}.
\newblock \emph{\bibinfo{journal}{Optics Express}}
  \textbf{\bibinfo{volume}{23}}, \bibinfo{pages}{10481--10497}
  (\bibinfo{year}{2015}).

\bibitem{Vaccari2014}
\bibinfo{author}{Vaccari, A.} \emph{et~al.}
\newblock \bibinfo{title}{{Light-opals interaction modeling by direct numerical
  solution of Maxwell’s equations}}.
\newblock \emph{\bibinfo{journal}{Optics Express}}
  \textbf{\bibinfo{volume}{22}}, \bibinfo{pages}{27739--27749}
  (\bibinfo{year}{2014}).

\bibitem{Chen2001}
\bibinfo{author}{Chen, J.~K.} \& \bibinfo{author}{Beraun, J. . E.~.}
\newblock \bibinfo{title}{{Numerical Study of Ultrashort Laser Pulse
  Interactions with Metal Films}}.
\newblock \emph{\bibinfo{journal}{Numerical Heat Transfer}}
  \textbf{\bibinfo{volume}{40}}, \bibinfo{pages}{1--20} (\bibinfo{year}{2001}).

\bibitem{Chen2007}
\bibinfo{author}{Chen, F.}, \bibinfo{author}{Du, G.}, \bibinfo{author}{Yang,
  Q.}, \bibinfo{author}{Si, J.} \& \bibinfo{author}{Hou, H.}
\newblock \bibinfo{title}{{Ultrafast Heating Characteristics in Multi-Layer
  Metal Film Assembly Under Femtosecond Laser Pulses Irradiation. Two Phase
  Flow, Phase Change and Numerical Modeling}}  (\bibinfo{year}{2011}).

\bibitem{zhig2012}
\bibinfo{author}{Zhigilei, L.~V.} \& \bibinfo{author}{Lin, Z.}
\newblock \bibinfo{title}{{Electron-Phonon Coupling and Electron Heat Capacity
  in Metals at High Electron Temperatures}} (\bibinfo{year}{2012}).

\bibitem{kittel}
\bibinfo{author}{Kittel, C.}
\newblock \emph{\bibinfo{title}{{Introduction to Solid State Physics}}}
  (\bibinfo{publisher}{Wiley}, \bibinfo{year}{1996}), \bibinfo{edition}{7} edn.

\bibitem{Ren2012}
\bibinfo{author}{Ren, Y.}
\newblock \emph{\bibinfo{title}{{A Comprehensive Model for Energy Transport and
  Ablation of Metal Films Induced by Ultrashort Pulsed Lasers}}}.
\newblock Ph.D. thesis, \bibinfo{school}{Missouri-Columbia}
  (\bibinfo{year}{2012}).

\bibitem{Vial2008}
\bibinfo{author}{Vial, A.} \& \bibinfo{author}{Laroche, T.}
\newblock \bibinfo{title}{{Comparison of gold and silver dispersion laws
  suitable for FDTD simulations}}.
\newblock \emph{\bibinfo{journal}{Applied Physics B}}
  \textbf{\bibinfo{volume}{93}}, \bibinfo{pages}{139--143}
  (\bibinfo{year}{2008}).

\bibitem{Ahmmed2014}
\bibinfo{author}{Ahmmed, K.}, \bibinfo{author}{Grambow, C.} \&
  \bibinfo{author}{Kietzig, A.-M.}
\newblock \bibinfo{title}{{Fabrication of Micro/Nano Structures on Metals by
  Femtosecond Laser Micromachining}}.
\newblock \emph{\bibinfo{journal}{Micromachines}} \textbf{\bibinfo{volume}{5}},
  \bibinfo{pages}{1219--1253} (\bibinfo{year}{2014}).

\end{thebibliography}
\end{multicols}
\end{document}